%% file: 000-springer.tex
\RequirePackage{fix-cm}

\documentclass[smallextended]{svjour3}       

\smartqed  
\usepackage{amsmath}
\usepackage{amssymb}
\usepackage{mathptmx}
\usepackage{varioref}
\usepackage{wrapfig}
\usepackage{enumerate}
\usepackage{nomencl3}
\usepackage{subfigure}
\usepackage{subfigmat}
\usepackage{fancyvrb}

\usepackage{lettrine}
\usepackage{graphicx}
\usepackage{amsmath}
\usepackage{color}
\usepackage{epsfig}
\usepackage{float}
\usepackage{overcite}
\usepackage{ifthen}
\usepackage{url}
\usepackage{booktabs}
\usepackage{gensymb}
\usepackage{tikz}
\usepackage{trimclip}
\usepackage{longtable}
\usepackage{needspace}
\usepackage{array}

\DeclareMathOperator{\sign}{sgn}
\DeclareMathOperator{\derivd}{d}

\def\AR{\text{\itshape\clipbox{0pt 0pt .32em 0pt}\AE\kern-.30emR}}

\makeglossary

\renewcommand{\nomgroup}[1]{%
  \ifthenelse{\equal{#1}{A}}{}{%
    \ifthenelse{\equal{#1}{B}}{\item[\textbf{Subscripts}]}{\item[\textbf{Superscripts}]}}{}}

 \journalname{Theoretical and Computational Fluid Mechanics}
\begin{document}

\input 010-title.tex

\date{Received: date / Accepted: date}

\maketitle
\input 020-overview

\setcounter{page}{1}

\printglossary[0.8in]

\input 015-nomen.tex

\input 030-introduction.tex

\input 040-theoretical.tex
\input 080-results.tex
\input 090-summary.tex

\input 095-ack.tex

\bibliographystyle{spmpsci}      

\bibliography{JabRefDatabase,kiran_bibtot}   

\end{document}

%% file: 010-title.tex
\title{ Unsteady lifting-line theory and the influence of wake vorticity on aerodynamic loads}

\titlerunning{ULLT wake vorticity}

\author{Hugh J.A. Bird \and Kiran Ramesh}

\authorrunning{Hugh Bird and Kiran Ramesh}

\institute{Hugh J.A. Bird \at
           Aerospace Sciences Division, School of Engineering, \\       
           University of Glasgow, Glasgow, G12 8QQ, UK \\
           \email{hugh.bird.1@research.gla.ac.uk}
           \and
           Kiran Ramesh \at
           Aerospace Sciences Division, School of Engineering, \\       
           University of Glasgow, Glasgow, G12 8QQ, UK \\
           \email{kiran.ramesh@glasgow.ac.uk}      
}

%% file: 020-overview.tex
\keywords{Unsteady lifting-line theory \and vortex dynamics \and unsteady aerodynamics}

\begin{abstract}

Frequency domain Unsteady Lifting-Line Theory (ULLT) provides a means by which the aerodynamics of 
oscillating wings may be studied at low computational cost without neglecting the 
interacting effects of aspect ratio and oscillation frequency.

Renewed interest in the method has drawn attention to several uncertainties however.
Firstly, to what extent is ULLT practically useful for rectangular wings, despite
theoretical limitations? And secondly, to what extent is a complicated wake model needed 
in the outer solution for good accuracy?

This paper aims to answer these questions by presenting a complete ULLT based on the work
of Sclavounos, along with a novel ULLT that considers only the streamwise vorticity and a
Prandtl-like pseudosteady ULLT. These are compared to Euler CFD for
cases of rectangular wings at multiple aspect ratios and oscillation frequencies. 

The results of this work establish ULLT as a low computational 
cost model capable of accounting for interacting finite-wing and oscillation frequency effects,
and identify the aspect ratio and frequency regimes where the three ULLTs are most 
accurate. This research paves the way towards the construction of 
time-domain or numerical ULLTs which may be augmented to account for non-linearities such as flow separation.

\end{abstract}

%% file: 015-nomen.tex
\section{Nomenclature}

{\renewcommand\arraystretch{1.0}
\noindent\begin{longtable}{@{}l @{\quad=\quad} l@{}}
$A_0, A_1 ... A_N$ & Fourier coefficients in 2D bound vorticity distribution\\
$\AR$ & aspect ratio\\
$c$ & chord\\
$C_l$ & lift coefficient (2D)\\
$C_L$ & lift coefficient (3D)\\
$C_m$ & moment coefficient (2D)\\
$C_M$ & moment coefficient (3D)\\
$F$ & 3D correction strength\\
$h$ & plunge displacement\\
$h_0^*$ & normalized plunge amplitude\\
$k$ & chord reduced frequency\\
$K$ & 3D interaction kernel\\
$s$ & semispan\\
$t$ & time\\
$U$ & free stream velocity\\
$W$ & 2D downwash\\
$x$ & chordwise coordinate\\
$x_m^*$ & normalized moment reference location\\
$x_p^*$ & normalized pitch location\\
$y$ & spanwise coordinate\\
$\alpha$ & angle of attack\\
$\gamma$ & vorticity per unit area\\
$\Gamma$ & bound circulation per unit span\\
$\zeta$ & angular reference coordinate on span\\
$\nu$ & span reduced frequency\\
$\phi$ & velocity potential\\
$\omega$ & angular frequency\\
\end{longtable}}

%% file: 030-introduction.tex
\section{Introduction}
\label{sec:intro}
The unsteady aerodynamics of wings is an increasingly important topic
in modern aerospace research. On the smallest of scales, engineers are
using flapping wing motions in the design of micro air vehicles
(MAVs). At larger scales, oscillatory motion is seen as a means by
which energy can be captured from water currents with reduced
environmental damage\cite{Rostami2017}. And at larger scales still,
high aspect ratio aircraft\cite{Wang2010, Murua2012a} and wind turbine
designers\cite{Hansen2006} face challenges is understanding the
consequences of wings flexing under load. 

The focus of this article is on finite-wing effects in unsteady
aerodynamics. Early research in this topic was strongly motivated by
dynamic stall in helicopters~\cite{mccroskey}. Studies in this topic,
carried out using Computational Fluid Dynamics (CFD) and experiments
are reviewed in Carr~\cite{carr1988progress}, Carr et al.~\cite{carr2}
and Ekaterinaris and
Platzer~\cite{ekaterinaris1998computational}. More recent
investigations (in the 21st century) into three-dimensional effects in
dynamic stall using state-of-the-art experimental and numerical
methods include Angulo et al.~\cite{andreu2019influence} and Visbal and
Garmann~\cite{visbal2019dynamic,visbal2019effect}. All these studies
concern the forces and flow around a finite aspect ratio wing
undergoing pitch-plunge motion in a high Reynolds number, low reduced
frequency regime.  The modern aerospace applications listed earlier
have inspired unsteady aerodynamics research in new regimes, at
low/transitional Reynolds numbers and involving high reduced
frequencies. These flows are typically massively separated and involve
leading-edge vortex (LEV) formation and shedding. Finite-wing effects
in these regimes, specifically concerning the interactions of tip
vortices with leading- and trailing-edge vortices have been
investigated using CFD and experiments, for translating
wings~\cite{mulleners2017flow,
  mancini2015unsteady,devoria2017mechanism}, pitching
wings~\cite{jantzen2014vortex,hord2016leading,yilmaz2012flow,yilmaz2010scaling,visbal2017unsteady,green2011unsteady},
plunging
wings~\cite{calderon2013volumetric,calderon2014absence,calderon2013lift,visbal2013three,yilmaz2010three,fishman2017structure},
rotating
wings~\cite{ozen2012three,carr2013finite,carr2015aspect,medina2016leading,beals2015lift,venkata2013leading}
and wings subject to
gusts~\cite{perrotta2017unsteady,corkery2018development,biler2019experimental}. These
studies have shed much light on force production in unsteady
finite-wings and on the contributions from circulatory, apparent-mass
and vortical effects.

Theoretical or low-order methods in aerodynamics are required for
rapid design analysis and optimization, and, later on in the
engineering cycle, real-time simulation and use in control
systems. The development of these is aided and inspired by
experimental and computational studies, such as those referenced
earlier. In 2D, Theodorsen's theory~\cite{theodorsen1935} provides the
lift and moment for a thin, symmetric airfoil undergoing small
harmonic pitch and plunge oscillations in potential flow. This theory
has been shown to provide good results even in regimes outside the
expected limit of validity such as at low Reynolds numbers, for large
amplitudes of oscillation, and even when LEVs are present (so long as
they are attached)~\cite{mcgowan2011investigations,ol2009shallow}. 2D
low-order models based on the discrete-vortex method have been used to
successfully predict airfoils undergoing arbitrarily large kinematics
and massively-separated
flows~\cite{kiran_journal1,ramesh2014discrete}.

In 3D, the simplest approach is strip
theory~\cite{Leishman2006}. Neglecting all three dimensional effects
(for large aspect ratios) allows two dimensional methods to be applied
to the wing in strips. In instances where this is a poor
approximation, low-order 3D methods such as the unsteady
boundary-element method~\cite{moored2018unsteady}, unsteady vortex
lattice method~\cite{Katz2001, Murua2012a,Smyth2019,hirato2019vortex} or vortex
particle methods~\cite{willis2007combined} can be used, but at far
greater computational cost due to the large numbers of
interacting vortex elements required to form wake vortex structures. 
Eldredge and Darakananda \cite{Eldredge2015} suggest forming these
vortex structures with a very small number of time-varying elements instead, but
this introduces difficulties. As with similar 2D models, the lack
of natural vortex shedding mechanisms makes such an approach dynamically incomplete
for long-term simulations, according to Darakananda and Eldredge \cite{Darakananda2018}.





There is however an intermediate to 2D and 3D methods. By augmenting
2D models with corrections gathered from a much simplified 3D model,
lifting-line theory may be obtained. By avoiding full three
dimensional interaction, the computational complexity of the problem
is significantly reduced. Lifting-line theory assumes a high aspect
ratio wing. The chord scale and span scale are so far removed that the
problem can effectively be separated into detailed inner 2D solutions
which interact via a simplified outer 3D solution.

The simplest form of lifting-line theory assumes straight wings and
steady flow. Prandtl\cite{Prandtl1923} is credited with the earliest
lifting-line theory. This was formalized through matched asymptotic
expansions by Van Dyke\cite{Dyke1964} and extended to curved and swept
wings by Guermond\cite{Guermond1990}. Non-linear extensions to steady
lifting-line theory have become an important design
tool\cite{Gallay2015}.

The desire for an unsteady lifting-line theory (ULLT) lead to work on
the problem of oscillating wings. Such a problem is the
three-dimensional analogue of the oscillating plate, analyzed by
Theodorsen\cite{theodorsen1935}. Assuming a high aspect ratio wing,
Cheng~\cite{cheng1976lifting} has identified five frequency ranges
based on the wake wavelength $\lambda$, span ($2s$) and average chord
($\overline{c}$): (i) very low frequency ($2s<<\lambda$), (ii) low
frequency ($2s=\mathcal{O}(\lambda))$, (iii) intermediate frequency
($\overline{c} << \lambda << 2s$), (iv) high frequency
($\overline{c}=\mathcal{O}(\lambda)$), and very high frequencies
($\overline{c} >> \lambda$). Early authors in this topic such as
James\cite{James1975} and van Holten\cite{Holten1976} assumed a
uniform induced downwash over the chord (as in the steady case) for
all frequencies. Ahmadi and Widnall\cite{Ahmadi1985} showed that this
assumption was asymptotically valid only for very low frequencies, and
further showed that in the low-frequency regime the induced downwash
has a harmonic variation in the chordwise direction. Guermond and
Sellier\cite{Guermond1991} showed that this conclusion is actually
true for all frequencies, and also derived a theory applicable for
wings with curvature and sweep and uniformly valid for all
frequencies. Sclavounos\cite{Sclavounos1987} produced a ULLT that
assumes a uniform induced downwash but is claimed to be valid at all
but very high frequencies.

All these ULLTs assume harmonic kinematics, small amplitudes of
motion, planar wakes, and use Theodorsen's theory to represent the 2D
solution in chordwise strips along the span of the wing. However, unlike in 2D where
the frequency-domain solution (Theodorsen) has a time-domain
equivalent (Wagner's indicial response)\cite{garrick1938}, it has not
yet been possible to derive time-domain equivalents for ULLTs.
 

Low-order models based on lifting-line theory in the time domain hence
employ more assumptions and approximations in comparison with harmonic
ULLTs. A common assumption made is to treat the 2D strip-wise
solutions as unsteady (using nonlinear inner 2D methods which may
apply to non-harmonic kinematics), while treating the trailing
vortices behind the wing as steady (in essence, Prandtl's LLT applied
to a time-varying wing circulation distribution). This assumption is
referred to as the ``pseudosteady assumption'' in this article.

The first unsteady finite wing analysis in the time domain is due to
Jones\cite{Jones1939}, although it is only applicable to elliptic
planforms. A time-domain ULLT is presented by Boutet and
Dimitriatis\cite{Boutet2018} using an inner solution based on Wagner's
method\cite{Wagner1925}, and a pseudosteady assumption. Ramesh et
al.\cite{Ramesh2017} introduce a lifting-line theory for a
geometrically non-linear inner solution, again using a pseudosteady
assumption. Devinant\cite{Devinant1998} presents a method by which the
pseudosteady assumption can be numerically removed for wake
wavelengths of the wing span scale or larger. Bird et
al.\cite{Bird2019} apply this method to a geometrically non-linear
inner solution. Sugar-Gabor et al.\cite{Sugar-Gabor2018} suggested a
method where the unsteady wake was only accounted for in the outer 3D
problem, allowing for roll and yaw kinematics. 
Lifting-line theory has also been used to model wings
in large-scale simulations where resolving the chord-scale would
be prohibitively expensive, for example in Caprace et al. \cite{Caprace2020}.


Strip theory (no 3D effects) and the pseudosteady assumption (steady
3D effects through Prandtl's LLT) are attractive owing to their
mathematical simplicity and easy of implementation. However, the
implications of the pseudosteady wake model - or any other simplified
wake model - are not well understood. Whilst Guermond and Sellier's
analysis \cite{Guermond1991} found that a pseudosteady wake model was
only asymptotically valid in the very low frequency regime, the
practical implications of this assumption in aerodynamic flows are
less clear.


Additionally, only limited validation of unsteady lifting-line
theory has been undertaken. 
 Sclavounos \cite{Sclavounos1987} and Boutet and Dimitriatis \cite{Boutet2018} 
compared their methods to the unsteady vortex lattice method. This method shares
some assumptions with unsteady lifting-line theory, and requires care in discretization to satisfy the Kutta condition~\cite{Roesler2018}. Sclavounos compared results against the whole-wing lift coefficient
of a rectangular and an elliptic wing undergoing oscillating kinematics. Boutet and Dimitriatis examined whole-wing lift and moments
for indicial and oscillating kinematics.
Guermond and Sellier \cite{Guermond1991} compared their work to a 
panel method for a pointy wing.
Much of experimental and CFD work
undertaken on oscillating wings (listed earlier) post-dates
publication of unsteady lifting-line theories. 
These works show that
the vortical topology of the wake is far more complex than might be
assumed, even at relatively low amplitudes.

Therefore, this paper firstly investigates the accuracy of a ULLT based
upon Sclavounos' interaction kernel
in comparison to CFD for rectangular wings oscillating in heave
and pitch. The use of Euler CFD avoids the explicit assumptions of
unsteady vortex lattice methods used in previous research.
Whilst lifting-line theory is not strictly valid in the
context of rectangular wings \cite{Dyke1964}, they are of great practical
relevance. 

Secondly, this paper modifies this `complete' ULLT in order to investigate
the effects of simplified wake models and the accuracy of frequency domain
unsteady lifting-line theory. 
The work is based upon a framework similar
to that of Sclavounos
because of its arrangement as the solution of an integro-differential
equation (akin to Prandtl's LLT), unlike the work of Van
Dyke\cite{Dyke1964} or Guermond and Sellier\cite{Guermond1991}. This
also makes it more amenable to modification of the 3D interaction
kernel. Understanding the trade-off between wake-model
complexity and ULLT accuracy is useful for those building more complex
numerical ULLTs.

The article is therefore laid out as follows:
in section~\ref{sec:theory}, ULLT is introduced,
based upon the framework of Sclavounos with
minor modifications. The 3D
interaction kernel and alternative forms of the kernel representing
different wake assumptions are discussed in section~\ref{sec:kernel}. 
The `complete' ULLT is validated against CFD in heave and pitch 
for the Euler regime in section~\ref{sec:cl_cullt}.
Next, the impact of using simplified ULLTs is explored for
whole wing loads in section~\ref{sec:_ULLT_kernel_result_comparison}, 
the wake vorticity distribution in section~\ref{sec:_ULLT_kernel_result_comparison_wake_vort},
and wing load distributions in section~\ref{sec:_lift_and_moment_dists}.
A brief discussion of the results obtained using the simplified wakes
is then given in section~\ref{sec:_wake_model_choice_advice}.
Section~\ref{sec:exp_val} compares
the ULLTs to experimental data, confirming that ULLTs are relevant 
even when the flow is not inviscid
before the conclusions are finally presented in section~\ref{sec: conclusions}.

%% file: 040-theoretical.tex
\section{Theoretical approach}
\label{sec:theory}


Sclavounos' ULLT considers potential flow past a thin, unswept wing
subject to a uniform freestream velocity ($U$) and undergoing
small-amplitude, harmonic pitch ($\alpha$) and plunge ($h$)
oscillations. The problem is formulated with an outer and inner
domain. At distances from the wing that are large compared to its chord
(outer domain), the flow is insensitive to the wing geometry. The
perturbation due to the wing may hence be approximated by a line of
concentrated bound circulation $\Gamma(y)$ along the $y$-axis and a
planar trailing vortex sheet representing the shed wake. At distances
from the wing comparable to the average chord (inner domain), the
problem is reduced to a sequence of airfoil sections undergoing
small-amplitude oscillations in pitch and plunge (Theodorsen's
problem). The final unsteady lifting-line solution is obtained by
matching asymptotic expansions of the inner and outer solutions.

In section~\ref{sec:outer}, Sclavounos' derivation of the outer solution
is reproduced and the kernel representing the unsteady induced
downwash is introduced. In section~\ref{sec:inner}, the inner solution
from Theodorsen's theory is presented in terms of a general unsteady
thin-airfoil formulation. Matching of the inner and outer asymptotic expansions and the final lifting-line equation (from Sclavounos) are presented in section~\ref{sec:matching}. Finally, the kernel and the influence of the
various wake models on the same are discussed in
section~\ref{sec:kernel}.

\subsection{The outer solution}
\label{sec:outer}

\begin{figure}
  \centering
  \includegraphics[width=0.98\textwidth]{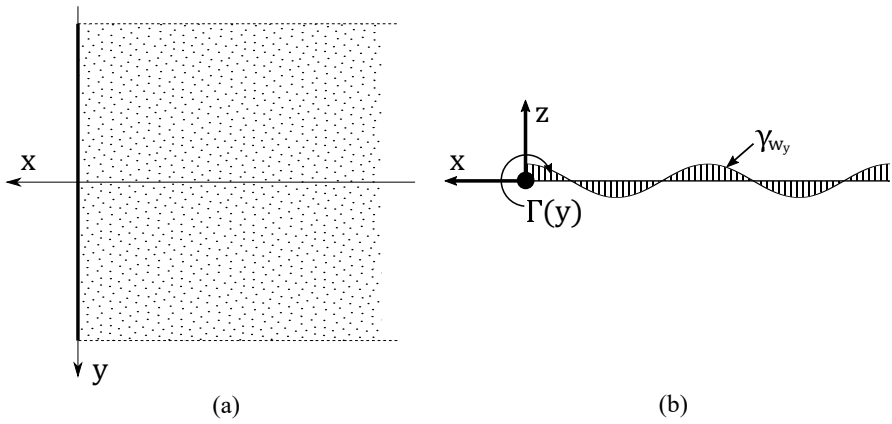}
  \caption{(a) Plan view of the outer domain consisting of the lifting
    line and its wake, and (b) view of the 2D problem in the outer
    domain.}
  \label{fig:outer}
\end{figure}

The lifting line and its oscillating wake as seen in the outer domain
are illustrated in figure~\ref{fig:outer}.

The bound vorticity / circulation of the wing is described by the
complex quantity,
\begin{equation}
  \Gamma (y,t) = \Gamma_{0}(y)e^{i \omega t}, \qquad |y| \leq s.
\end{equation}

Kelvin's theorem dictates that the change
in bound circulation is convected into the wake. Hence the spanwise ($\gamma_{w_y}$) and
streamwise ($\gamma_{w_x}$)  wake vorticity at the wing trailing-edge are

\begin{align}	
	\gamma_{w_y}|_{x=0} &= -\frac{1}{U}\frac{\partial \Gamma}{\partial t} = -\frac{i \omega \Gamma_{0}(y)}{U}e^{i \omega t} \label{eq: kelvin condition spanwise} \\	
	\gamma_{w_x}|_{x=0} &= \frac{\partial \Gamma_{0}(y)}{\partial y}e^{i \omega t} \label{eq: kelvin condition streamwise}
\end{align}

Vorticity is convected only by the free stream, hence
\begin{equation}
	\label{eq: vorticity sinusoidal convection}
	\gamma_{w_{x/y}}(x, y, t) = 
	\begin{cases}
	  \gamma_{w_{x/y}}(0, y, t) e^{i \omega x/U}, & x \leq 0\\
	  0, & x > 0
	\end{cases}
\end{equation}

Now consider a two dimensional slice in the $x$-$z$ plane, as
shown in figure~1b. From this 
perspective, the lifting-line appears as a point vortex, with a wake
on $z=0$, $x \leq 0$.

It is in this plane that the Kelvin condition for spanwise circulation,
given by equation~\ref{eq: kelvin condition spanwise}, is
satisfied. For a bound circulation of unit amplitude, the velocity
potential is given by
\begin{equation}
	\overline{\phi^{2D}}(x, z) = \frac{1}{2\pi}\tan^{-1}{\frac{z}{x}} - \frac{i \omega}{2\pi U} \int_{-\infty}^{0}  e^{i \omega \xi/U} \tan^{-1}\frac{z}{x - \xi} \derivd \xi
\end{equation}

The first term represents the bound circulation at origin and the second
term represents the 2D wake vortex sheet.

The streamwise vorticity, $\gamma_{w_x}$, due to equation~\ref{eq: kelvin
  condition streamwise} must still be included.  The effects of this
are given as the second term in the asymptotic expansion of the full
outer velocity potential equation, derived by Sclavounos as

\begin{equation}
	\phi(x, y, z, t) \thicksim \Gamma(y,t) \overline{\phi^{2D}}(x, z) - \frac{z}{2\pi} \int^s_{-s} \frac{\partial \Gamma(\eta,t)}{\partial\eta} K(y - \eta) \derivd \eta
\label{eqn:outer_exp}
\end{equation}

\noindent where the kernel $K(y)$ will be discussed in section~\ref{sec:kernel}. This
second term represents an unsteady downwash accounting for the 3D
interaction in the outer domain.  Differentiating this term with
respect to $z$ gives the induced downwash due to finite wing
effects. Since this second term is invariant with respect to $x$ and
linear in $z$, the downwash at any spanwise location is uniform in the
$x$-$z$ plane over the section chord. This is a simplification that
restricts the asymptotic validity of the theory to span-scale wake
wavelengths and longer (very-low and low frequencies) according to
Guermond and Sellier\cite{Guermond1991}. The effect of this assumption
in predicting loads on the wing are investigated and discussed in
section~\ref{sec: results} by comparing lift coefficients from ULLT against
numerical simulations over a large frequency range.

\subsection{The inner solution}
\label{sec:inner}
In the inner domain, the detail of the flow around a chord section is
considered. Since it is assumed that $2s\gg c$, and that changes in the
flow happen on the span length-scale, the problem can be considered as
2D. The problem consists of a symmetric airfoil at a certain spanwise
section undergoing small-amplitude pitch and plunge oscillations,

\begin{align}
  \begin{split}
  h(y,t) &= h^*_0(y)c(y)e^{i \omega t} \\ 
  \alpha(y,t) &= \alpha_0(y)e^{i (\omega t + \psi)} 
\label{eqn:kinem}
  \end{split}
\end{align}

\noindent where $h^*_0$ is plunge amplitude per unit chord, $\alpha_0$
is pitch amplitude, $\psi$ is the phase between plunge and pitch, and
$\omega$ is the frequency of oscillation typically expressed as a
chordwise or spanwise reduced frequency

\begin{equation}
  k(y) = \frac{\omega c(y)}{2U}, \qquad \nu = \frac{\omega s}{U}
\end{equation}

\begin{figure}
  \centering
   \includegraphics[width=0.6\textwidth]{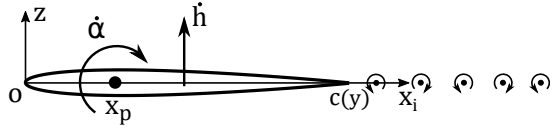}
   \caption{Inner domain, showing the velocities of the airfoil,
    pitch-axis location, chord length and spanwise shed vorticity.}
  \label{fig:inner}
\end{figure}

Figure~\ref{fig:inner} shows this problem, for which the solution is
given by Theodorsen\cite{theodorsen1935}. The results are expressed
here in terms of a general unsteady thin-airfoil
theory\cite{Katz2001,kiran_journal1} which allows easy
modification of the approach for various wing geometries and
kinematics. The bound vorticity distribution over the airfoil section
is taken as a Fourier series

\begin{equation}
  \gamma(x,y,t) = 2U\left[A_{0}(y,t)\frac{1+\cos \theta}{\sin \theta}
  + \sum_{n=1}^{\infty}A_{n}(y,t) \sin(n\theta)\right]
\label{eqn:tat_vort}
\end{equation}

\noindent where $\theta$ is a transformation variable related to the
chordwise coordinate as $x_i=0.5c(1-\cos\theta)$, and $A_{0}$,
$A_{1}$...$A_{N}$ are time-dependent Fourier coefficients that vary
over the span. The $A_0$ term represents the suction peak caused by
the flow having turn around the airfoil leading edge, and the Kutta
condition (zero vorticity at the trailing-edge) is enforced implicitly
through the form of the Fourier series. The Fourier coefficients are
determined by enforcing the zero-normal-flow boundary condition on the
aerofoil chord line.

The bound airfoil circulation and the sectional lift and pitching
moment coefficient from unsteady thin-airfoil theory are given by

\begin{align}
  \Gamma(y,t) &= Uc\pi \left(A_0 + \frac{A_1}{2} \right) \label{eqn:gamma} \\
  C_l(y,t) &= 2 \pi \Biggl[ A_0 + \frac{1}{2}A_1 + \frac{c}{U}\left(\frac{3}{4}\dot{A_0} + \frac{1}{4}\dot{A_1} +  \frac{1}{8}\dot{A_2}\right)\Biggl]  \label{eqn:cl}\\
  \begin{split}
    C_m(y,t) &= 2 \pi \Biggl[ A_0 \left(x_m^* - \frac{1}{4}\right) +\frac{A_1}{2}\left(x_m^* - \frac{1}{2}\right) + \frac{A_2}{8} \\
       &+ \frac{c}{U}\left( \frac{3\dot{A_0}}{4}\left(x_m^* - \frac{7}{12}\right) + \frac{\dot{A_1}}{4}\left(x_m^* - \frac{11}{16}\right) + \frac{\dot{A_2}}{8}\left(x_m^*-\frac{1}{2}\right) + \frac{\dot{A_3}}{64}\right)\Biggl]
  \end{split}
  \label{eqn:cm}
\end{align}

\noindent where $x_m^*$ is the reference location per chord (ranging
from $0-1$) about which the pitching moment is calculated.

Ramesh\cite{ramesh2020leading} has derived the Fourier coefficients
corresponding to Theodorsen's solution as

\begin{align}
  A_0 &= C(k) \frac{W_{3qc}}{U} - \frac{\dot{\alpha}c}{4U} \nonumber \\ 
  A_1 &= \frac{\dot{\alpha}c}{2U} - 2\frac{W_{3qc}}{U}(C(k) - e^{-ik}S(k)) \nonumber \\ 
  A_{2 ... N} &= (-1)^n 2ikS(k)\frac{W_{3qc}}{U}Q_n 
  \label{eqn:fourier_theo}
\end{align}

\noindent where $C(k)$ is Theodorsen's function, 

\begin{equation}
C(k) = \frac{K_1(ik)}{K_1(ik) + K_0(ik)}
\end{equation}

\noindent and S(k) is the Sears function,

\begin{equation}
  S(k) = \frac{1/ik}{K_1(ik) + K_0(ik)}
\end{equation}

\noindent where $K_0(z)$ and $K_1(z)$ are modified Bessel functions
of the second kind \cite{Olver2010}.
 $Q_n(k)$ are wake coefficients,

\begin{equation}
  Q_n(k) = \int_0^\infty e^{-ik \cosh \zeta} e^{-n \zeta} d \zeta
\end{equation}

\noindent and $W_{3qc}$ is the normal downwash at the airfoil
three-quarter chord location, calculated from the kinematics as

\begin{equation}
  W_{3qc} = U \alpha - \dot{h} - \dot{\alpha}c\left(x_p^*-\frac{3}{4}\right)
  \label{eqn:downwash}
\end{equation}

Using equations~\ref{eqn:kinem} and \ref{eqn:downwash}, the 2D normal
downwash at three-quarter chord location at any spanwise location from
plunge and pitch are

\begin{align}
  W^{2D}_{3qc, h}(y,t) &= - 2iUk h^*_0(y) e^{i \omega t} 
  \label{eqn:downwash_theo_heave}\\
  W^{2D}_{3qc, \alpha}(y,t) &= U\Biggl[1 - 2ik \left(x_p^*-\frac{3}{4}\right)\Biggl] \alpha_0(y)e^{i (\omega t + \psi)} 
  \label{eqn:downwash_theo_pitch}
\end{align}

The 2D bound circulation, lift coefficient and moment coefficient for
pitch and plunge kinematics at any spanwise location are calculated from
equations~\ref{eqn:gamma}, \ref{eqn:cl}, \ref{eqn:cm}, \ref{eqn:fourier_theo}, \ref{eqn:downwash_theo_heave} and \ref{eqn:downwash_theo_pitch} as

\begin{align}
  \Gamma^{2D}_{h}(y,t) &= \frac{4 U h^*_0(y)c(y) e^{-ik}}{i H_0^{(2)}(k) + H_1^{(2)}(k)}e^{i \omega t} \\
  \Gamma^{2D}_{\alpha}(y,t) &= \frac{4U\alpha_0(y)c(y) e^{-ik}}{i H_0^{(2)}(k) + H_1^{(2)}(k)}\left(\left(x_p^*-\frac{3}{4}\right) -  \frac{1}{2 i k}\right)e^{i (\omega t+\psi)} \\
  C_{l_h}(y,t) &= 2\pi h^*_0 (-2ikC(k) + k^2 ) e^{i \omega t} \label{eq:2d_C_l_h}\\
  C_{l_{\alpha}}(y,t) &= 2\pi\alpha_0\left[C(k)\left(1-2ik\left(x^*_p-\frac{3}{4}\right)\right)+\frac{ik}{2}+k^2\left(x^*_p-\frac{1}{2}\right)\right]e^{i (\omega t + \psi)} \\ 
   C_{m_h}(y,t) &= 2\pi h^*_0 \left[ - 2ik C(k) \left(x^*_m-\frac{1}{4}\right) + k^2\left(x^*_m-\frac{1}{2}\right)\right] e^{i \omega t}   \label{eq:2d_C_m_h} \\
      C_{m_{\alpha}}(y,t) &= 2\pi\alpha_0\Bigg[C(k)\left(1-2ik\left(x^*_p-\frac{3}{4}\right)\right)\left(x^*_m-\frac{1}{4}\right) \nonumber \\
       &+ k^2\left(x^*_p\left(x^*_m-\frac{1}{2}\right) - \frac{1}{2}\left(x^*_m-\frac{9}{16}\right)\right)+\frac{ik}{2}\left(x^*_m-\frac{3}{4}\right)\Bigg] e^{i (\omega t + \psi)}
 \end{align}
\noindent where $H_0^{(2)}(z)$ and $H_1^{(2)}(z)$ are Hankel 
functions of the second kind, and $x_p^*$ is the pivot location.

\subsection{Matching of the inner and outer solutions}
\label{sec:matching}

The asymptotic inner expansion of the outer solution velocity
potential is given in equation~\ref{eqn:outer_exp}. For the inner
solution, the velocity potential is constructed as a combination of
homogeneous and particular solutions~\cite{Sclavounos1987}. The
homogeneous solution (to model 3D induced effects) is determined from
the interaction of the airfoil with a uniform downwash, and the
particular solution is obtained from the 2D solutions for pitch and
plunge kinematics derived above.

\begin{equation}
  \phi(x,y,z,t) = \phi^{2D}_{h/\alpha} + F(y)(i\omega z e^{i \omega t} - \phi^{2D}_{hn})
\end{equation}

\noindent where the first and second terms are the particular and
homogeneous solutions, respectively. The uniform downwash $i\omega z 
e^{i \omega t}$ corresponds to unit plunge amplitude $h_0=h_0^*c=1$ for
which the associated velocity potential is $\phi^{2D}_{hn}$ with
resulting bound circulation $\Gamma_{hn}^{2D}$.
The outer expansion of the inner velocity potential at large distances from
the chord is given as

\begin{equation}
  \phi(x,y,z,t)  \thicksim \Gamma^{2D}_{h/\alpha}\overline{\phi^{2D}} + F(y)(i \omega z e^{i \omega t} - \Gamma^{2D}_{hn}\overline{\phi^{2D}})
  \label{eqn:inner_exp}
\end{equation}


The outer and inner expansions, equations~\ref{eqn:outer_exp}
and~\ref{eqn:inner_exp}, may now be matched. Equating the
$\overline{\phi_{2D}}$ terms allows an expression for the bound
circulation distribution to be extracted as
\begin{equation}
	\Gamma(y,t) = \Gamma^{2D}_{h/\alpha}(y,t) - F(y)\Gamma^{2D}_{hn}(t)
        \label{eqn:match1}
\end{equation}

\noindent leaving 

\begin{equation}
	F(y) = -\frac{1}{2 \pi i \omega e^{i \omega t}} \int_{-s}^s \Gamma'(\eta)K(y - \eta)\derivd \eta
        \label{eqn:match2}
\end{equation}

The contributions to circulation from pitch and plunge may be added
together since the theory is linear. The lifting-line integro-differential equation for
the time-varying spanwise circulation is obtained from equations~\ref{eqn:match1}
and~\ref{eqn:match2} as

\begin{equation}
  \label{eq: integro-diff equation}
  \Gamma - \frac{\Gamma^{2D}_{hn}}{2 \pi i \omega} \int^s_{-s} \Gamma'(\eta)K(y-\eta) \derivd\eta = \Gamma^{2D}_{h} + \Gamma^{2D}_{\alpha}
\end{equation}

We note that all terms in the lifting-line equation contain the common
factor $e^{i \omega t}$. An approximate solution for the complex
circulation amplitude $\Gamma_0$(y) can be obtained by expressing it in a
Fourier series

\begin{equation}
  \Gamma_0 = 4Us \sum_{m=1}^M \Gamma_{m} \sin (m\zeta)
\end{equation}

\noindent where $y = -s \cos\zeta$. For problems where both the
kinematics and planform of the wing are symmetric about $y=0$, the
even $m$ terms can be neglected. For a rectangular wing, the problem is then reduced to
numerically solving the equation,

\begin{multline}
	 \sum_{m=1}^{M} \Gamma_{m} \sin (m\zeta)  - \\
	 \frac{c e^{-ik}}{i\pi k (i H^{(2)}_{0}(k) + H^{(2)}_1(k))}\int^\pi_{0} \sum_{m=1}^M m \Gamma_{m} \cos(m\sigma) K(s \cos \sigma - s \cos\zeta) \derivd\sigma\\
	  \qquad = \frac{e^{-ik}}{\AR(i H^{(2)}_0(k) + H^{(2)}_1(k))}\left(2 h^*_0 + \alpha_0 e^{i \psi} \left(2\left(a-\frac{3}{4}\right)-\frac{1}{ik}\right)\right)      
\end{multline}

\noindent at collocation points distributed over the span. The 3D
induced downwash is given by

\begin{equation}
  W_i(y,t) = \frac{2U s e^{i \omega t}}{\pi}\int^\pi_{0} \sum_{m=1}^M m \Gamma_{m}\cos(m\sigma) K(s \cos \sigma - s \cos\zeta) \derivd\sigma
\end{equation}

The Fourier solutions $A_{0,1...N}(y,t)$ representing the inner solutions
across the span may then be corrected for 3D effects by evaluating
equation~\ref{eqn:fourier_theo} with the total 3D downwash at the
three-quarter chord location

\begin{equation}
  W^{3D}_{3qc}(y,t) = W^{2D}_{3qc}(y,t) + W_i(y,t)
\end{equation}

The allows the calculation of the stripwise circulation, lift
coefficient and pitching moment (accounting for 3D effects) from
equations~\ref{eqn:gamma},~\ref{eqn:cl} and~\ref{eqn:cm}. The wing lift and pitching moment are

\begin{equation}
  C_L = \frac{1}{s \overline{c}}\int_{0}^{s} C_l(y)c(y)\derivd y, \qquad C_M = \frac{1}{s \overline{c}^2}\int_{0}^{s} C_m(y)c^2(y)\derivd y 
\end{equation}

\noindent where the 2D coefficients have been corrected to include the 3D correction as $C_{l/m} = C_{l/m}^{2D} - F C_{l/m_{hn}}$.

\subsection{The kernel $K(y)$}
\label{sec:kernel}

The kernel $K(y)$, first introduced in equation~\ref{eqn:outer_exp},
represents the spanwise interaction of the inner solutions. It must
account for the difference between the inner solution, that assumes
that the flow is 2D, and the actual 3D nature of the problem via the
outer solution. The kernels based on various underlying assumptions
considered in this article are expressed below in terms of the
non-dimensional spanwise distance $y^*=y/s$. 

For strip theory, all three dimensional interaction is neglected. The
spanwise wake vorticity $\gamma_{w_y}$ is modeled in the inner
solution, but the strip theory outer solution neglects both the shed
streamwise vorticity $\gamma_{w_x}$, and the correction for the fact that
the variation of $\gamma_{w_y}$ over the span of the wing is not captured
by the inner solution. Since all interaction between the inner domains
is neglected, the strip theory kernel $K_{2D}$ is

\begin{equation}
	\label{eq: K strip theory}
	K_{2D}(y) = 0
\end{equation}

A pseudosteady kernel accounts for the shed streamwise vorticity
$\gamma_{w_x}$ in the outer domain, as given by equation~\ref{eq:
  kelvin condition streamwise}. However, it neglects the sinusoidal
variation with respect to downstream coordinate given in equation~\ref{eq: vorticity
  sinusoidal convection}. Again, the variation in $\gamma_{w_y}$ with
respect to span is not corrected for. The resultant pseudosteady
kernel $K_{P}$ is equivalent to that of Prandtl:

\begin{equation}
	\label{eq: K pseudosteady}
	K_{P}(y) = \frac{1}{2sy^*}
\end{equation} 

\noindent and the ULLT based on this kernel is abbreviated P-ULLT.

If the sinusoidal variation of $\gamma_{w_x}$ with respect to $x$
given by equation~\ref{eq: vorticity sinusoidal convection} is
accounted for the streamwise vorticity kernel $K_x$ can be
obtained. The Biot-Savart law can be applied to the streamwise
vorticity field. The downwash on a point of the
wing at $y_0$ due to a section of the wing $\derivd y$ is therefore
\[
	\derivd q = -\frac{\partial \Gamma}{\partial y}\frac{1}{4 \pi}
        \int^0_{-\infty}\frac{e^{\frac{i \omega \xi}{U}}(y-y_0)}{(x^2
          + (y - y_0)^2)^\frac{3}{2}} \derivd \xi
\]
allowing $K_S$ to be obtained as
\begin{equation}
	\label{eq: K x filament}
	K_{S}(y) = \frac{1}{2sy^*}\left[\nu|y^*| K_1\left(\nu|y^*|\right) +
          \frac{i\pi \nu |y^*|}{2}\Big(I_1\left(\nu|y^*|\right) -
          L_{-1}\left(\nu|y^*|\right)\Big)\right]
\end{equation}
where $I_n(x)$ and $K_n(x)$ are the modified Bessel functions of the
first and second kind respectively, and $L_n(x)$ is the modified
Struve function\cite{Olver2010}. Once again this neglects variation in
$\gamma_{w_y}$ with repect to span in the outer solution. The ULLT based
on this kernel is referred to the simplified ULLT (S-ULLT). 

Sclavounos obtained a kernel $K_C$ that accounts for both the shed
streamwise vorticity $\gamma_{w_x}$ and the 3D correction to the effects
of the shed spanwise vorticity $\gamma_{w_y}$.
\begin{equation}
  \label{eq: K unsteady}
  K_{C}(y) = \frac{1}{2s}\sign(y^*)\left[\frac{e^{-\nu |y^*|}}{|y^*|} - i\nu E_1(\nu|y^*|) + \nu P(\nu|y^*|) \right]
\end{equation}
where $E_1(x)$ is the exponential integral\cite{Olver2010} and 
\begin{equation}
  P(y) = \int^\infty_1 e^{-yt}\left[ \frac{\sqrt{t^2 - 1} - t}{t}\right]\derivd t  + i \int^1_0 e^{-yt}\left[\frac{\sqrt{1-t^2}-1}{t}\right] \derivd t
\end{equation}

The ULLT using Sclavounos' full kernel is denoted as the complete ULLT
(C-ULLT) in this research.

The different ULLT models described above may be summarized in terms
of the way the wake is modeled in the outer solution as shown in
table~\ref{tab:wake}.

\begin{table}
  \centering
  \caption{Features of the trailing wake behind the wing (in the outer
    solution) for the various solution methods considered in this
    research.}
\begin{tabular}{c c  l}
  \toprule			
  Method & Kernel & Wake model in outer solution \\
  \midrule
  Strip theory & $K_{2D}$ & $\gamma_{w_y}$: No model  \\
   & & $\gamma_{w_x}$: No model  \\
  \midrule
  P-ULLT  & $K_{P}$ & $\gamma_{w_y}$: No model  \\
  (Pseudosteady) & & $\gamma_{w_x}$: constant with respect to $x$-coordinate \\
  \midrule
  S-ULLT & $K_S$ & $\gamma_{w_y}$: No model  \\
  (Simplified) & & $\gamma_{w_x}$: harmonic variation in $x$-direction  \\
  \midrule
  C-ULLT & $K_C$ & $\gamma_{w_y}$: harmonic variation in $x$-direction  \\
  (Complete)& & $\gamma_w{_x}$: harmonic variation in $x$-direction  \\  
  \bottomrule
\end{tabular}
  \label{tab:wake}
\end{table}

At low frequencies the unsteady solution approaches the
pseudosteady solution. Accordingly
\begin{equation}
	\lim_{\omega \rightarrow 0}\{ K_C(y)\} = \lim_{\omega \rightarrow 0} \{ K_S(y)\}   = K_{P}(y) 
\end{equation} 

As the oscillation frequency tends to infinity, 3D effects become
negligible, and the kernels that include sinusoidal variation of the
shed wake approach the strip theory solution:
\begin{equation}
	\lim_{\omega \rightarrow \infty}\{ K_C(y)\} = \lim_{\omega \rightarrow \infty} \{ K_S(y)\}   = K_{2D}(y) = 0
\end{equation}

%% file: 080-results.tex
\section{Results}
\label{sec: results}

ULLT is derived from potential-flow theory based on incompressible
flow with zero viscosity, and employs further simplifying assumptions
of high aspect ratio and low reduced frequency. To determine the range
of validity of ULLT across the relevant range of aspect ratio and
reduced frequency, and to study the influence of the simplified wake
models on the solution, the ULLT is first compared against
numerical computations of the
incompressible Euler equations. In section~\ref{sec:exp_val}, the ULLT
models are validated against previously published experimental data in
realistic aerodynamic conditions. Comparative remarks about the ULLTs
are then made in section~\ref{sec:_wake_model_choice_advice}.

\subsection{Analysis with Euler Computational Fluid Dynamics}
\label{sec:cfd_analysis}

Numerical computations of the unsteady incompressible Euler equations
are performed using the open-source CFD toolbox OpenFOAM. A
body-fitted, structured computational mesh is moved according to
prescribed plunge and pitch kinematics, and the time-dependent
governing equations are solved using a finite volume method. A
second-order backward implicit scheme is used to discretize the time
derivatives, and second-order limited Gaussian integration schemes are
used for the gradient, divergence and Laplacian terms. The pressure
implicit with splitting of operators (PISO) algorithm implements
pressure-velocity coupling. This in-house setup has previously been
used with the incompressible Navier-Stokes governing equations to
study leading-edge vortex shedding on finite
wings~\cite{bird2018theoretical,Bird2019}, and with the
incompressible Euler equations to verify unsteady potential flow
solutions for an airfoil~\cite{ramesh2020leading}.

In this section, we use Euler CFD solutions to validate the 3D
aerodynamic loads obtained from ULLT, and to study the influence of
ULLT wake models on the loads and load distributions. In the CFD
setup, inviscid flow is considered with kinematic viscosity set to
zero, and a slip boundary condition is employed for the moving wing
surface. A NACA0004 section is chosen to best match the theoretical
assumptions of thin section and Kutta condition at the trailing edge.

\begin{figure}
\centering
    	\includegraphics[width=0.48\textwidth]{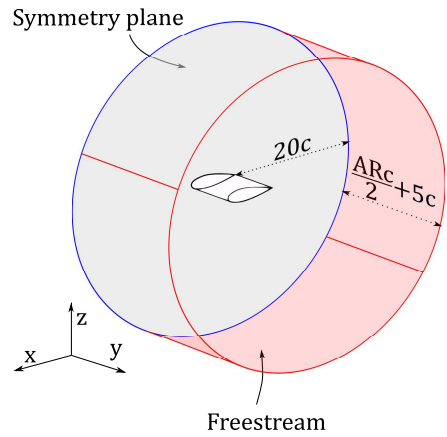}

	\caption{Dimensions of the mesh used for CFD. The symmetry boundary condition 
	is colored gray, and the freestream boundary condition is colored red.}
	\label{fig:cfd_domain_diagram}	
\end{figure}

Three aspect ratios (8, 4, and 2) are considered. Cylindrical O-meshes
for half the wings are constructed since the pitch and plunge
kinematics considered are all symmetric about the wing root. The
meshes have $160$ cells around the wing section, with increased
resolution near the leading and trailing edges. The wall-normal
direction has $115$ cells with the far-field extending $20$ chord
lengths in all directions around the section. In the spanwise
direction, the aspect ratio 8, 4 and 2 wings have $218$, $199$ and
$87$ cells respectively over the wing, with increased resolution
near the wingtip. For all three wings, the spanwise domain extends 5
chord lengths beyond the wingtip, with $100$ cells in this region. 
A diagram of the mesh is shown in Fig.~\ref{fig:cfd_domain_diagram}. Symmetry boundary condition is used for the circular domain at the wing root, and freestream (inlet/outlet) boundary conditions are used at the spanwise and wall-normal far-field domains. The freestream boundary condition behaves as a zero-gradient condition when fluid is flowing out of the boundary face, and as a fixed value condition (equal to freestream) when fluid is not flowing out.

Harmonic pitch and heave kinematics with chord reduced frequencies $k$
of 0.0, 0.125, 0.25, 0.5, 1.0 and 1.5 were simulated. Small amplitudes of
oscillation, $0.01c$ for plunge and $1\degree$ for pitch were used,
again to best satisfy the theoretical assumptions and ensure attached
flows. The CFD cases are shown in table~\ref{tab: CFD cases}.

\begin{table}
\centering
	\caption{CFD case specification for oscillating rectangular
          wings. All combinations of the below values were computed.}
	\label{tab: CFD cases}
	\begin{tabular}{l l}
	\toprule
	Property & Value\\
	\midrule
	Kinematics type & Heave, pitch about leading edge \\
	Wing planform & Rectangular\\
	Aspect ratio (\AR) & 2, 4, 8 \\
	Chord reduced frequency ($k$) & 0.0, 0.125, 0.25, 0.5, 1.0, 1.5\\
	\bottomrule
	\end{tabular}
\end{table}

\subsubsection{Validation of lift and moment coefficients from C-ULLT}
\label{sec:cl_cullt}

In this section the $C_L$ and $C_M$ (about mid-chord) obtained from Sclavounos' complete
ULLT are validated against the results obtained using CFD for heave
and pitch kinematics. For both heave and pitch, the loads are
normalized by the oscillation amplitude (since ULLT is linear with
respect to this parameter) and the wing has a zero mean pitch angle. For heave, the loads are further
normalized by the chord reduced frequency ($k$) since as seen in
equations~\ref{eq:2d_C_l_h} and \ref{eq:2d_C_m_h}, it is a common multiplier for $C_{l_h}$ and
$C_{m_h}$ (unlike for pitch). This is done to better differentiate the
various curves at low frequencies as seen below.

\vspace{10pt}\needspace{4\baselineskip}
\noindent\textbf{Heaving kinematics}

\begin{figure}
\centering
    \subfigure[]{
	\label{fig: rect heave abs CL}
    	\includegraphics[width=0.48\textwidth]{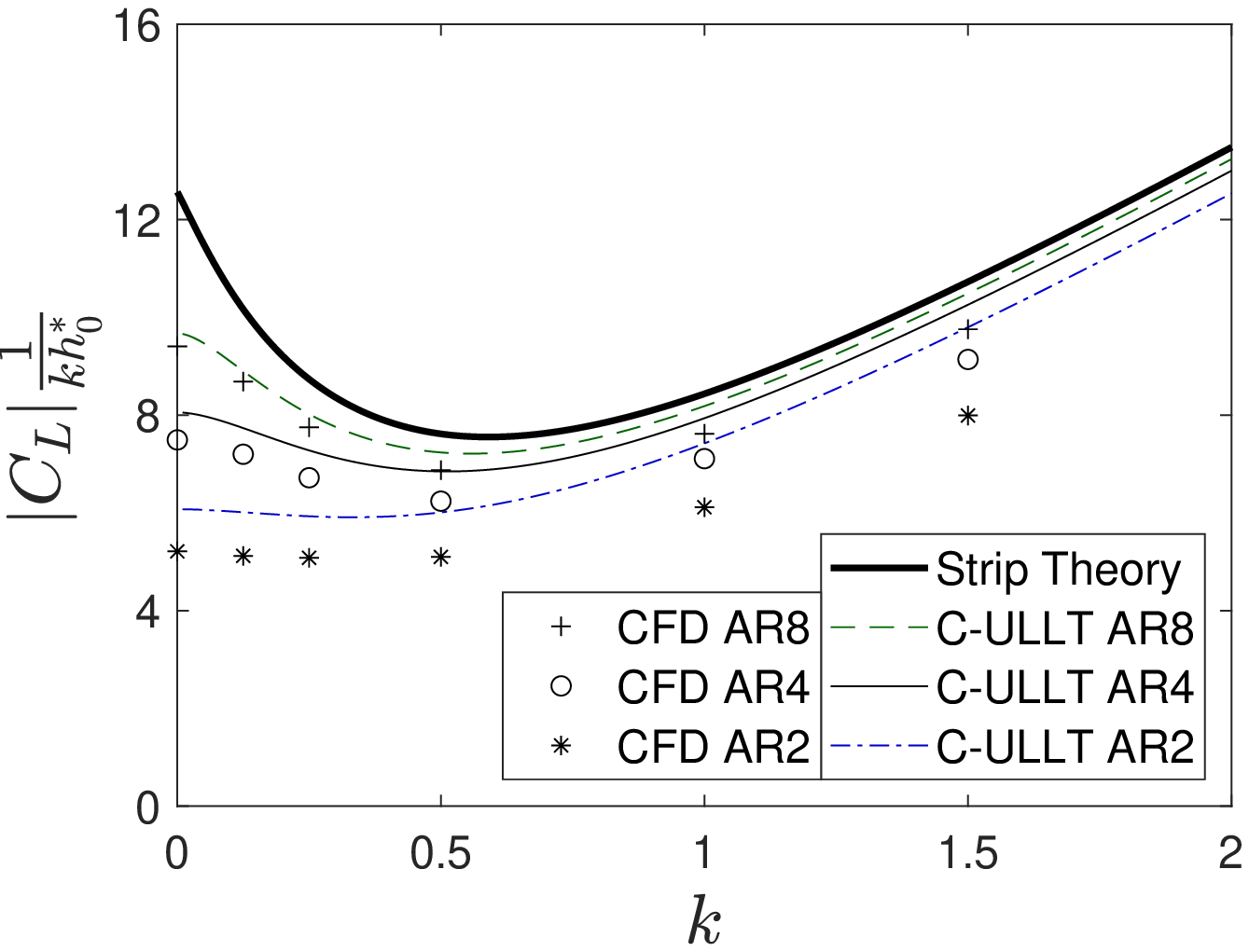}}
    \subfigure[]{
	\label{fig: rect heave ph CL}
    	\includegraphics[width=0.48\textwidth]{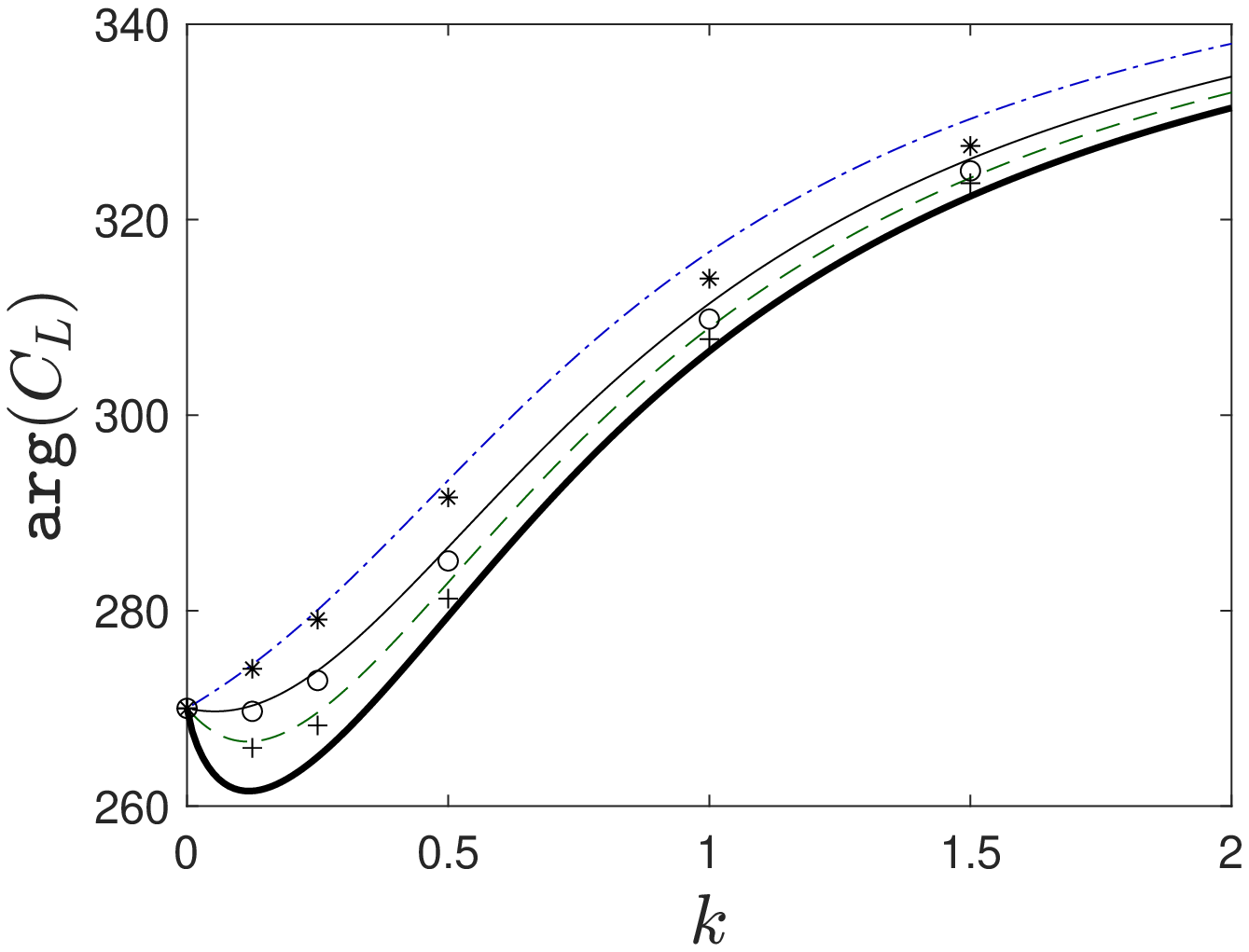}}

    \subfigure[]{
	\label{fig: rect heave abs CM}
    	\includegraphics[width=0.48\textwidth]{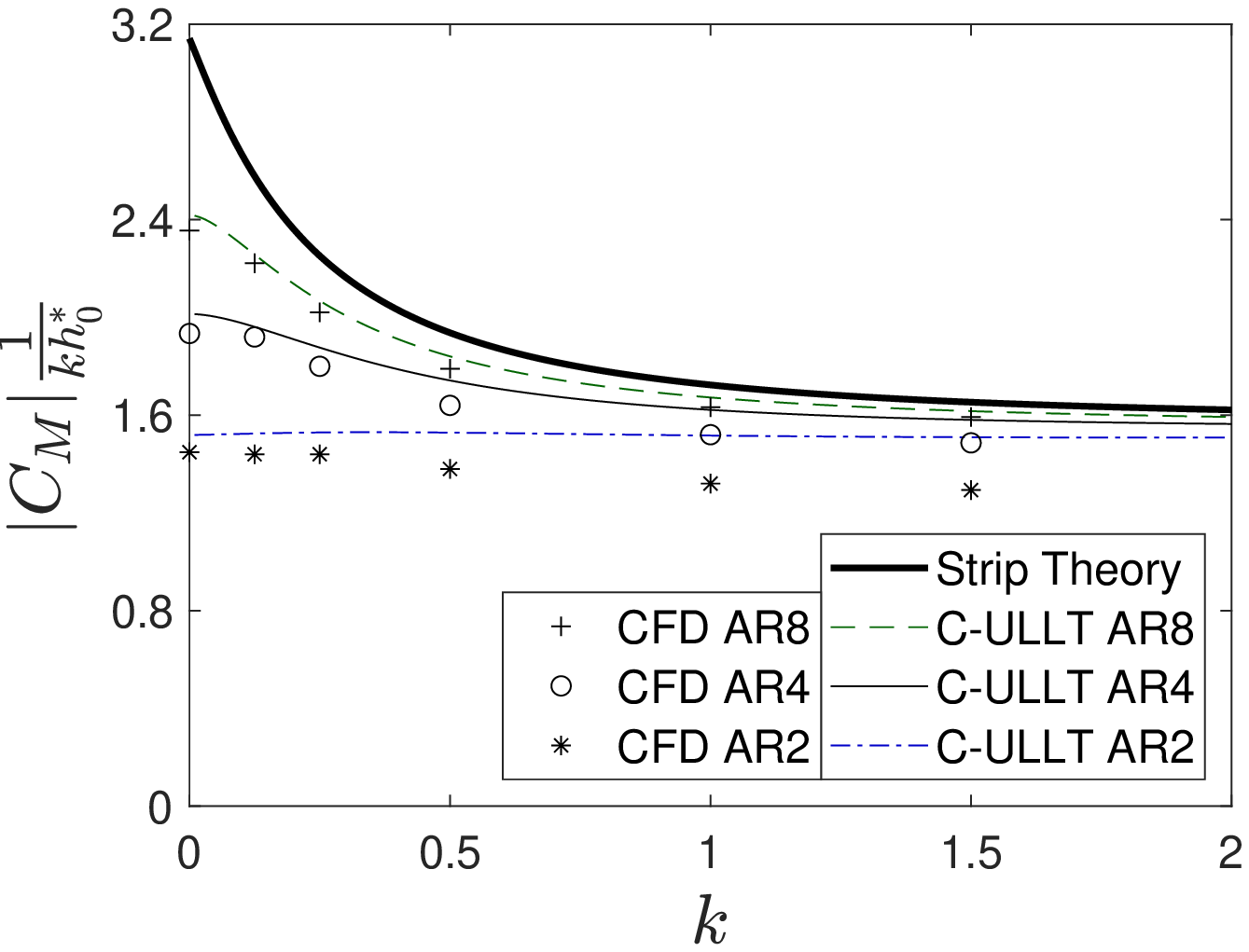}}
    \subfigure[]{
	\label{fig: rect heave ph CM}
    	\includegraphics[width=0.48\textwidth]{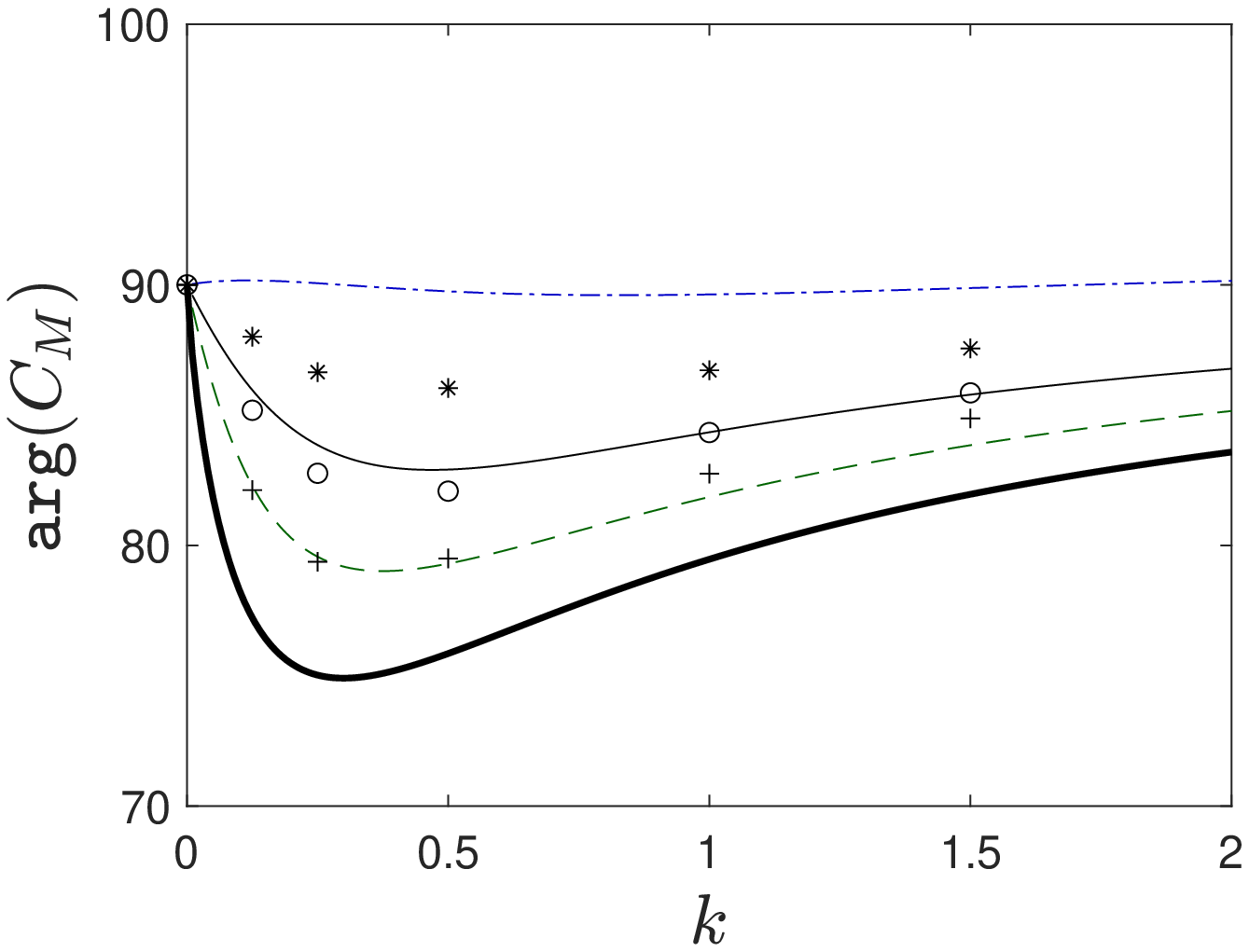}}

	\caption{Comparison of lift and moment about mid-chord for
          C-ULLT, strip theory and CFD for rectangular wings
          oscillating in heave.}
	\label{fig: heaving cfd comparison}	
\end{figure}

The normalized amplitude and phase of the wing lift and moment
coefficients for heave oscillations cases are shown in
figure~\ref{fig: heaving cfd comparison}. The theoretical results and
the CFD results are shown by lines and points respectively. Strip
theory (2D Theodorsen solution) is also shown by the bold line as a
reference; the differences between strip theory and the three ULLT
curves show the influence of 3D unsteady induced downwash. 

Examining the result from the CFD in figure~\ref{fig: heaving cfd
  comparison}(a), the normalized lift $|C_L|/ kh_0^*$ is lower for
lower aspect ratios across all frequencies. In the frequency range
approximately under $k=0.5$, $|C_L|/ kh_0^*$ decreases for \AR s 8 and 4,
and remains roughly constant for \AR 2. Above this frequency, the
$|C_L|/ kh_0^*$ curve increases for all \AR s, with the gradients of the
curves being slightly smaller for lower aspect ratios.

The C-ULLT (Sclavounos' complete ULLT) results in figure~\ref{fig:
  heaving cfd comparison}(a) broadly predict the trends found in
CFD. In the regime where ULLT assumptions are best satisfied, at high
aspect ratio 8 and for low frequencies, the comparison with CFD is
excellent. As aspect ratio decreases and as frequency increases, the
prediction worsens. At low frequencies, the errors in C-ULLT
predictions for \AR 4 are relatively small but larger for \AR 2. As
frequency increases, C-ULLT follows the trends of the curves from CFD,
but overpredicts $|C_L|/ kh_0^*$. there is also an overestimation of
the gradient, suggesting that added mass is overpredicted by
C-ULLT. This error increases as aspect ratio becomes smaller. 
 It is likely that this error is caused by the ULLT assumption
of 2D flow near the wing tips. This assumption is broken for 
rectangular and elliptic wings.

The phase of wing lift coefficient from CFD and C-ULLT are compared in
figure~\ref{fig: heaving cfd comparison}(b) . In the limit of low
frequency, the C-ULLT predicts that all aspect ratios would have the
same $C_L$ phase lag of $90\degree$, but that the response of the
phase with increasing $k$ would be different according to aspect
ratio. Higher aspect ratios increase the phase lag, whilst lower
aspect ratios decrease it. These trends agree well with the CFD
results. As frequency increases, the phase prediction obtained for
higher aspect ratios remains good but an offset is seen for the \AR 2 case. 

The comparison of normalized pitching moment $|C_M|/ kh_0^*$ between
C-ULLT and CFD is shown in figure~\ref{fig: heaving cfd
  comparison}(c). Similar to lift, the predictions are best for the
  high aspect ratio 8 wing. As \AR{} decreases, the error increases,
  particularly at high frequencies. In both CFD and C-ULLT, the
  $|C_M|/ kh_0^*$ curve approaches a limiting value with increasing
  $k$. However, contrary to the CFD results, the C-ULLT predicts that
  this value is independent of aspect ratio.
  
For the phase of $C_M$ shown in figure~\ref{fig: heaving cfd
  comparison}(d), the CFD shows that at all aspect ratios, the phase lead
  of $C_M$ initially decreases with $k$.  This phase lead then starts
  increasing again in the region of $k=0.5$. C-ULLT predicts the trends
  of the CFD correctly, particularly in the expected regime of
  validity (high aspect ratio, $k \approx < 1$).

Overall, the C-ULLT predicts the results and trends of the CFD
reasonably well, albeit worse at low aspect ratio. Importantly
however, even at such lower aspect ratios, it always provided a better
prediction of the CFD result than could be obtained using strip
theory.

\vspace{10pt}\needspace{4\baselineskip}
\noindent\textbf{Pitching kinematics}

\begin{figure}
	\centering
    \subfigure[]{
	\label{fig: rect pitch abs CL}
    	\includegraphics[width=0.48\textwidth]{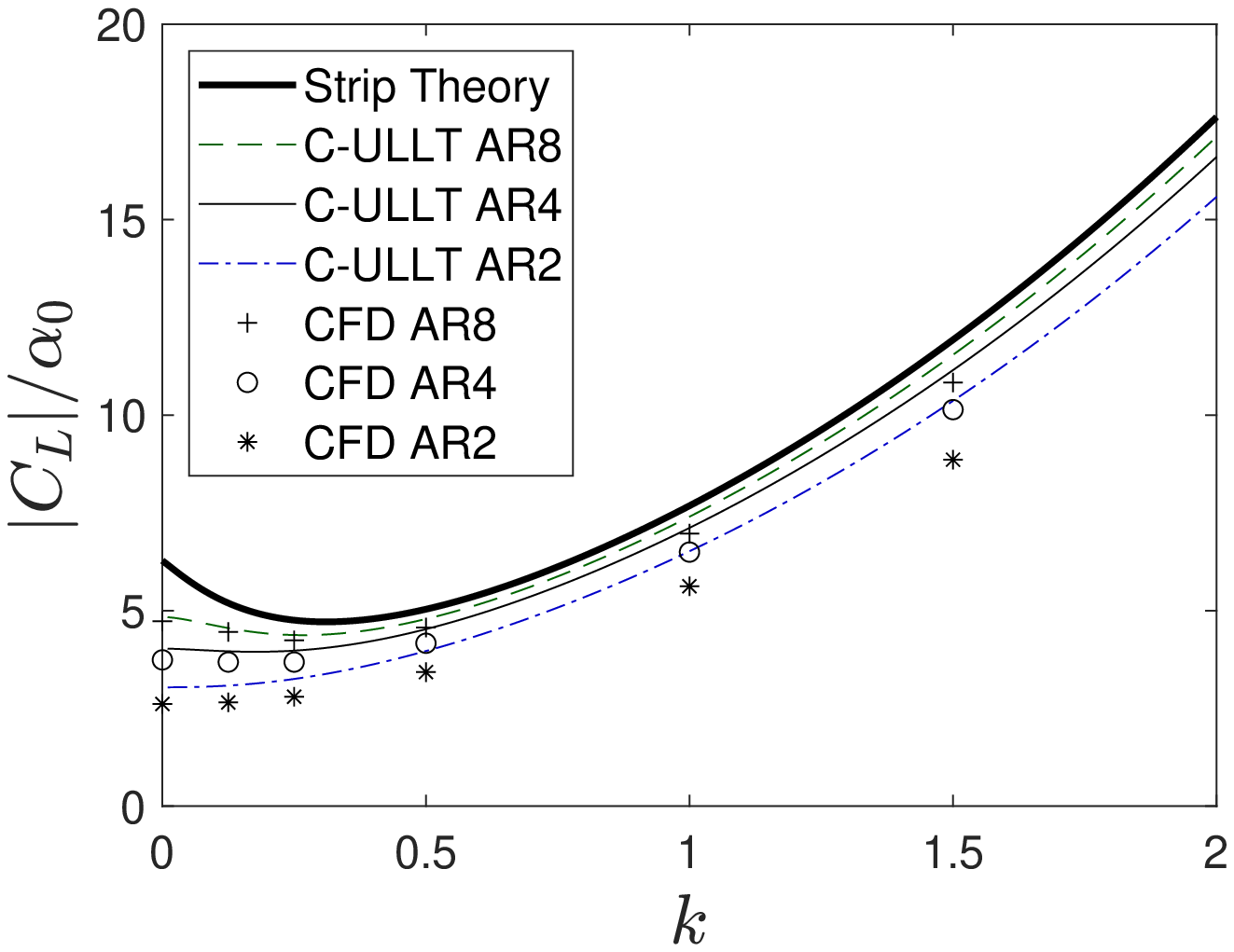}}
    \subfigure[]{
	\label{fig: rect pitch ph CL}
    	\includegraphics[width=0.48\textwidth]{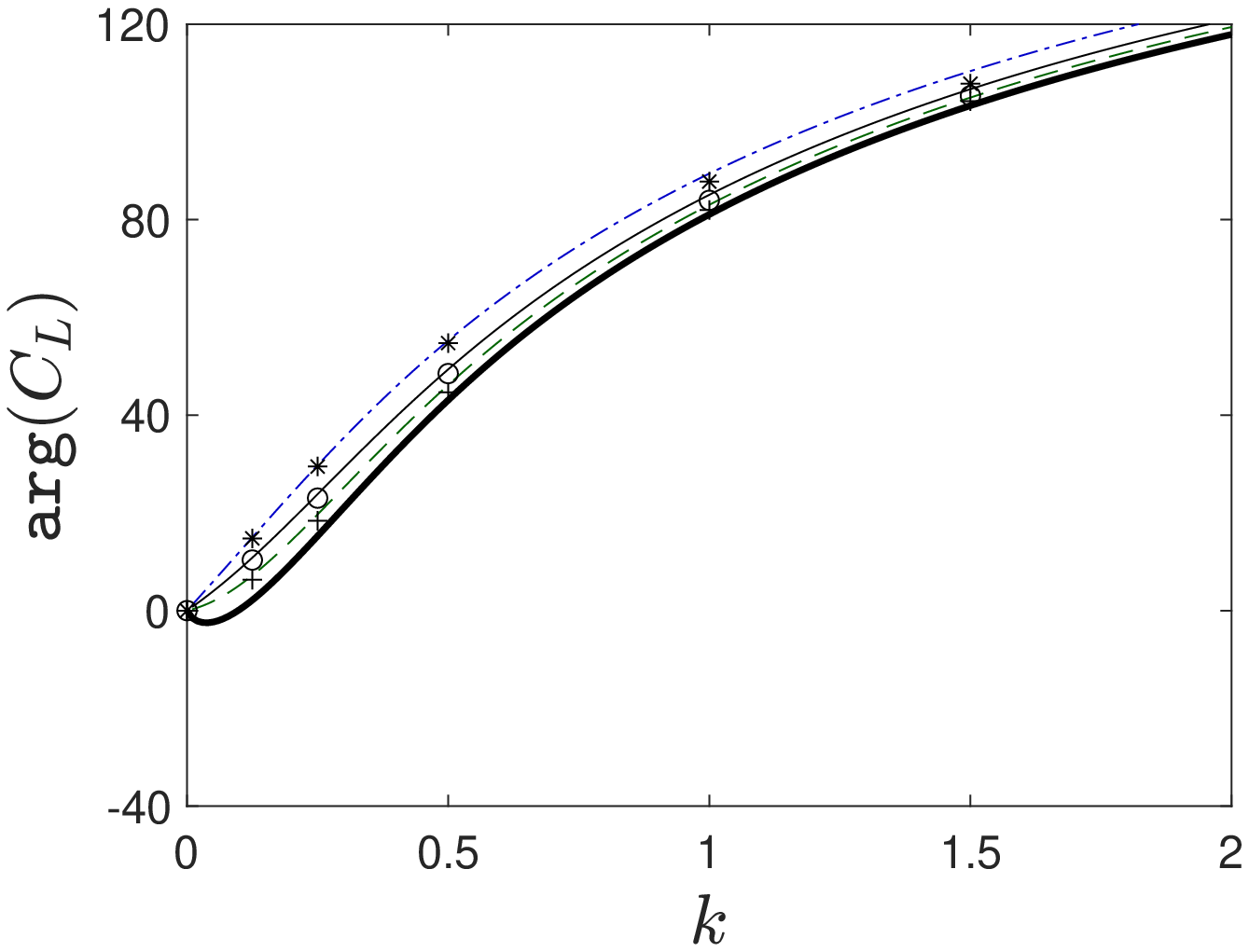}}

    \subfigure[]{
	\label{fig: rect pitch abs CM}
    	\includegraphics[width=0.48\textwidth]{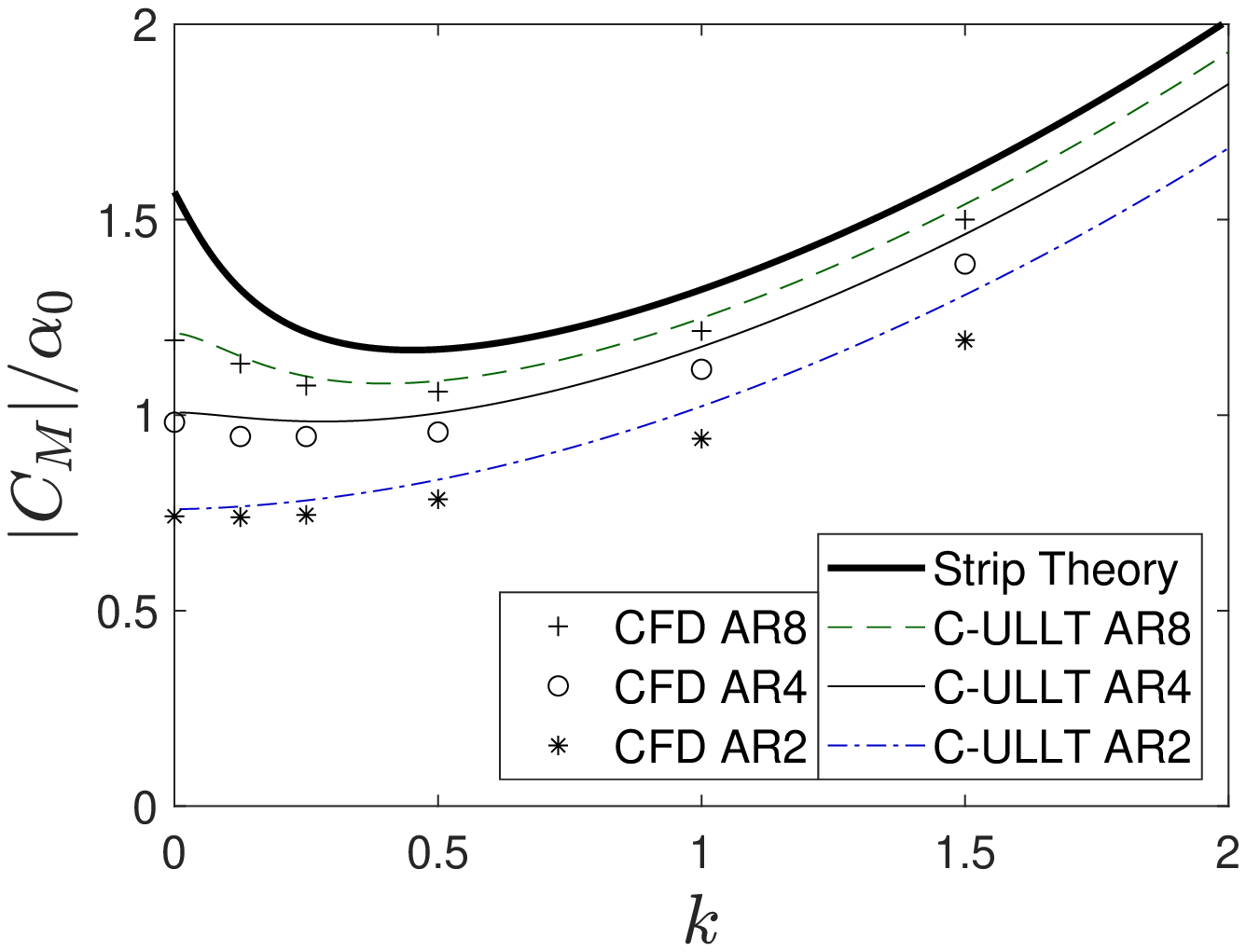}}
    \subfigure[]{
	\label{fig: rect pitch ph CM}
    	\includegraphics[width=0.48\textwidth]{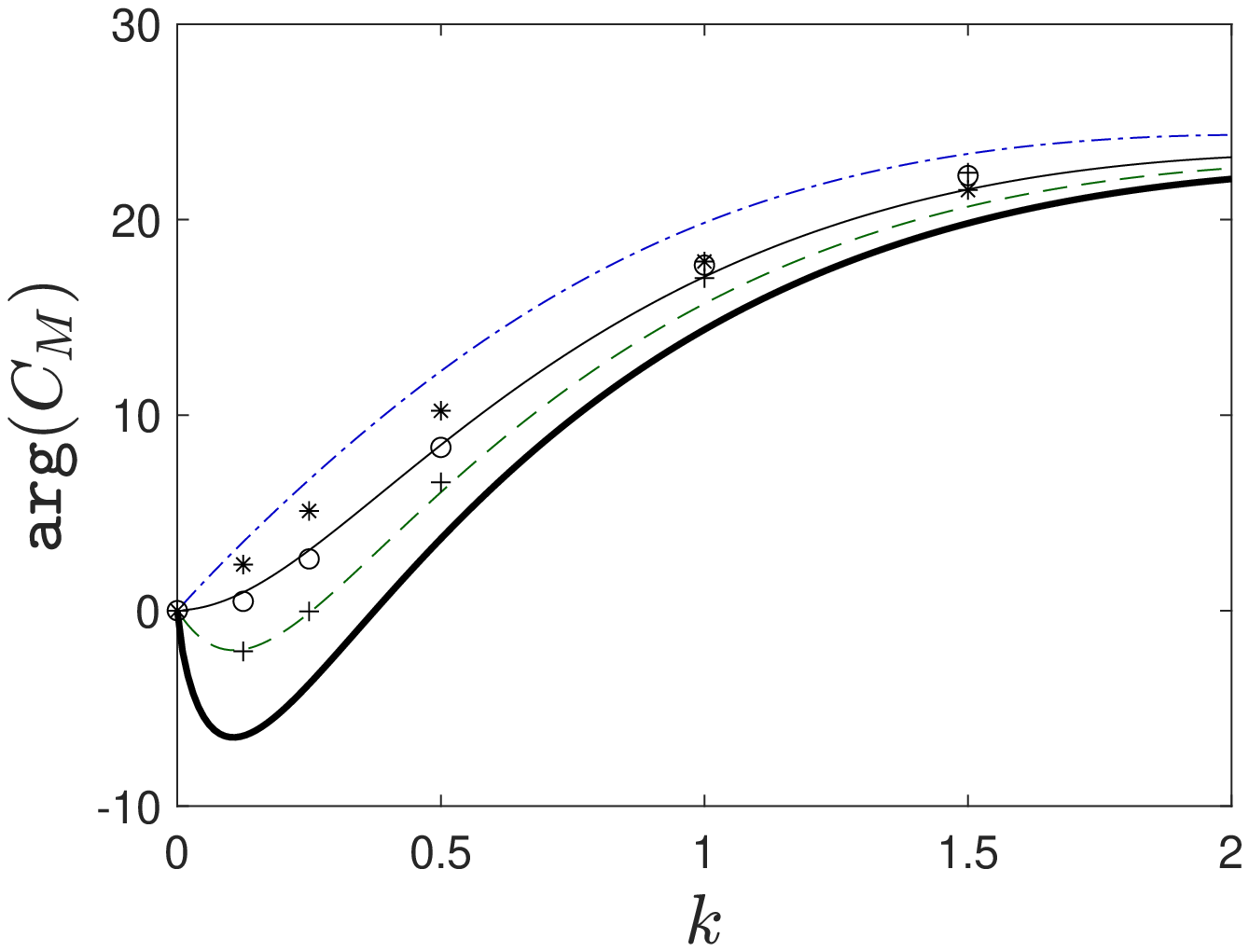}}

	\caption{Comparison of the C-ULLT, strip theory and CFD for a
          rectangular wings oscillating in pitch about the leading
          edge.  Moment coefficient is taken about the mid-chord.}
	\label{fig: pitching cfd comparison}	
\end{figure}

Next, the C-ULLT is validated against CFD for
leading-edge pitching kinematics. The amplitude and phase
of lift and moment coefficients for the pitch oscillations cases are
shown in figure~\ref{fig: pitching cfd comparison}.
 
The CFD result for the normalized lift amplitude $|C_L|/\alpha_0$ in
figure~\ref{fig: pitching cfd comparison}(a) initially decreases with
respect to $k$ at aspect ratio 8, stays constant at aspect ratio 4 and
increases slightly at aspect ratio 2 (until about $k \approx 0.25$).
Lower aspect ratio leads to a lower $|C_L|/\alpha_0$. As $k$ increases further,
$|C_L|/\alpha_0$ increases super-linearly. C-ULLT reflects the trends of the
CFD, with the initial gradient and y-intercept being dependent upon
aspect ratio. C-ULLT obtains a better prediction for all frequencies
with increasing aspect ratio, with errors increasing with higher
frequencies.

The CFD data in figure~\ref{fig: pitching cfd comparison}(b) shows
that initially the $C_L$ starts in phase to the kinematics. As the
frequency increases a phase lead develops. Lower aspect ratios result
in a larger phase lead. The phase of $C_L$ is excellently predicted by
C-ULLT at aspect ratio 8.  The accuracy of the prediction worsens at
lower aspect ratio, but it remains good, qualitatively reflecting the
trends of the CFD.

CFD results for the normalized pitching moment amplitude in
figure~\ref{fig: pitching cfd comparison}(c) show that $|C_M|/\alpha_0$
is lower for lower aspect ratios. C-ULLT correctly predicts the trends
of the CFD. For high aspect ratio, it constantly over-predicts the CFD
result by a small amount. At lower aspect ratios, the over-prediction
is initially small before growing.

In figure~\ref{fig: pitching cfd comparison}(d), CFD shows that $C_M$ is
initially approximately in phase with the kinematics. As frequency
increases, the $C_M$ leads the kinematics. At low frequencies,
the higher aspect ratios have lower phase lead. As frequency increase
the phases become more similar, approximately converging at $k\approx1$. As
frequency increases further, the higher aspect ratios develop larger
phase lead than the lower aspect ratios cases. C-ULLT provides a good
prediction of the aspect ratio 8 and 4 $\texttt{arg}(C_M)$ results at
approximately $k \approx\leq 0.5$. As frequency increases, the prediction of
C-ULLT worsens. Contrary to the CFD result, it predicts that the phase
lead of $C_M$ always decreases with aspect ratio.

Overall, it was found that C-ULLT was capable of predicting the CFD
results with good accuracy for rectangular wing pitching about the
leading edge. The C-ULLT always provided a better prediction of the
CFD results than strip theory, and was most accurate for \AR s 4 and 8
with $k \approx < 0.5$. 

\subsubsection{Influence of ULLT kernel on lift coefficient}
\label{sec:_ULLT_kernel_result_comparison}

\begin{figure}
\centering
    \subfigure[Aspect ratio 8]{
	\label{fig:_ar8_ULLT_comp}
    	\includegraphics[width=0.48\textwidth]{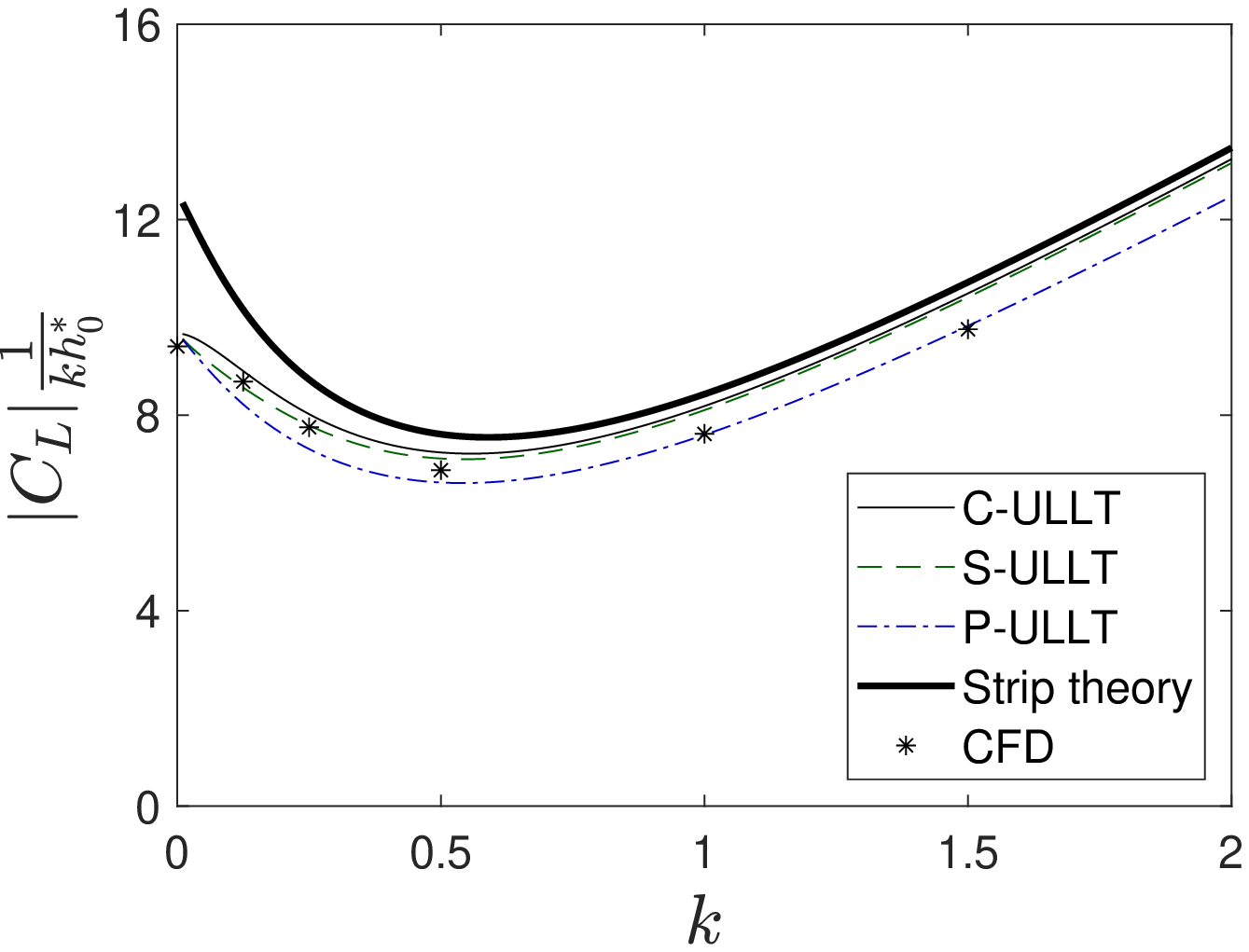}}
    \subfigure[Aspect ratio 4]{
	\label{fig:_ar4_ULLT_comp}
    	\includegraphics[width=0.48\textwidth]{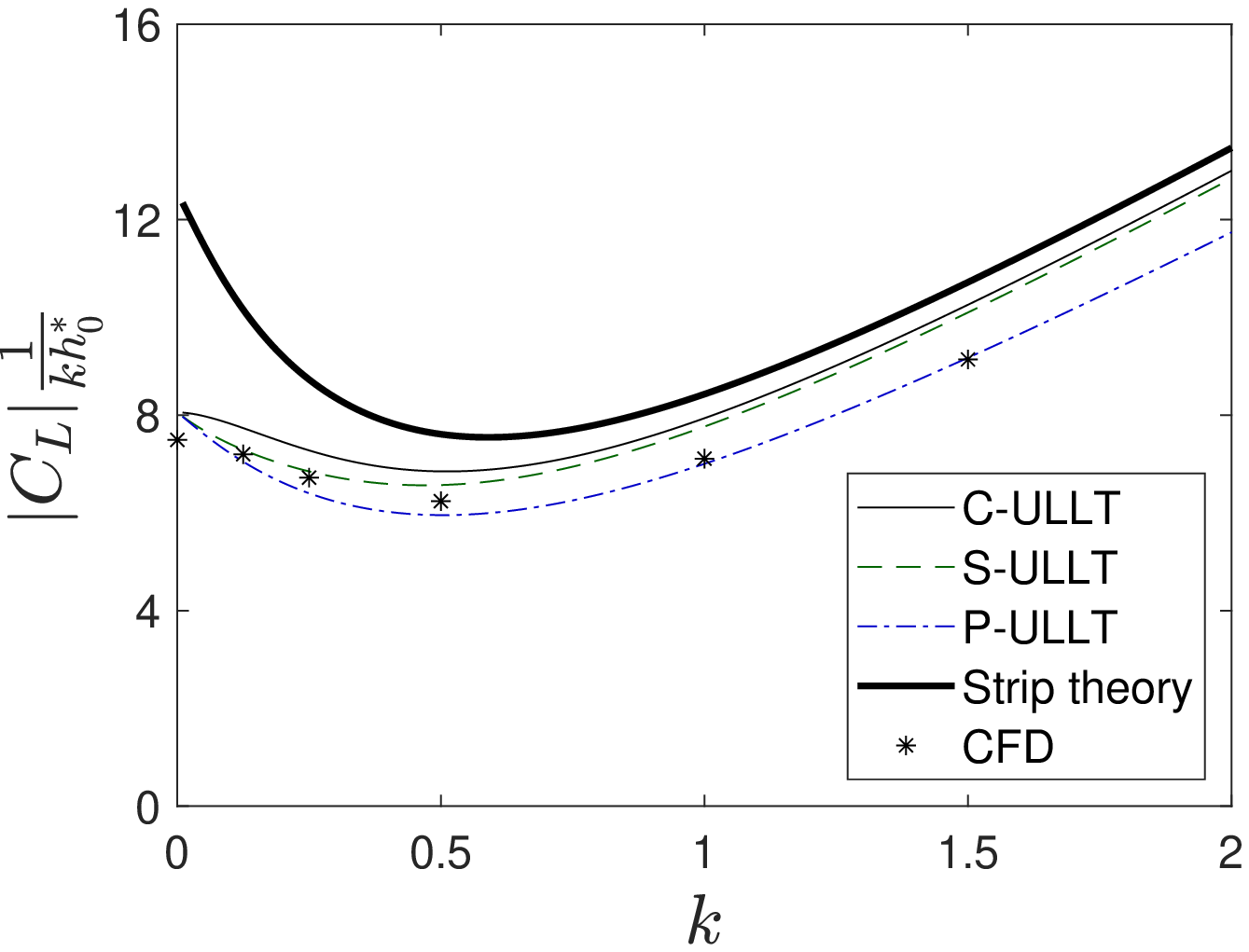}}

    \subfigure[Aspect ratio 2]{
	\label{fig:_ar2_ULLT_comp}
    	\includegraphics[width=0.48\textwidth]{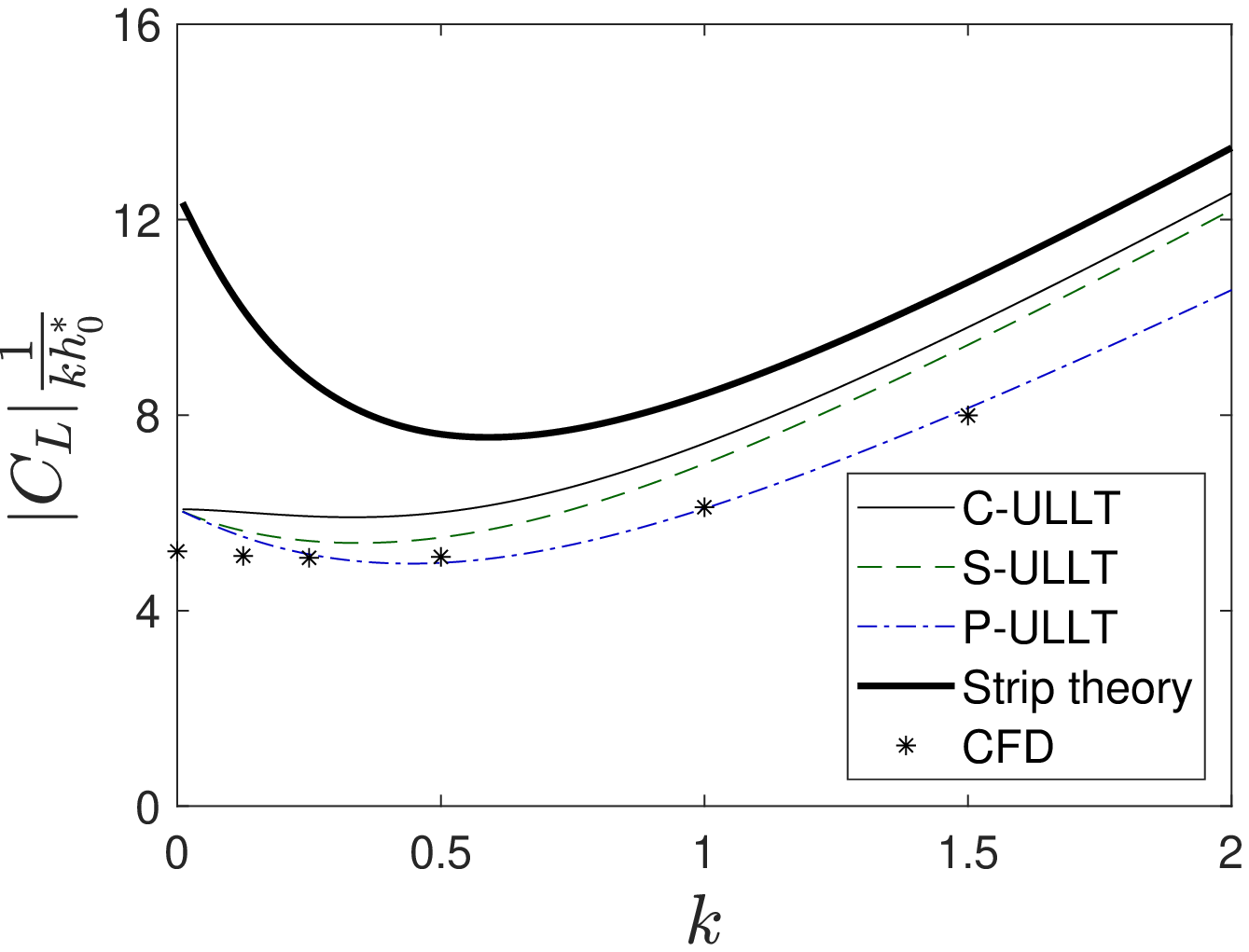}}

	\caption{Comparison of lift prediction from unsteady
          lifting-line theories for rectangular wings oscillating in
          heave.}
	\label{fig:_kernel_C_L_comparison}	
\end{figure}

In section~\ref{sec:cl_cullt} it was found that the agreement between CFD
and Sclavounos' complete ULLT was reasonably good for both lift and
moment coefficients, for both pitching and heaving kinematics. The
agreement was excellent in the ``ideal'' regime (for ULLT) of high
aspect ratio and low chord reduced frequency. In section~\ref{sec:kernel}
alternate ULLT models with approximated kernels representing the 3D
unsteady induced downwash were introduced. These were the S-ULLT
(simplified) and the P-ULLT (pseudosteady), whose assumptions are
summarized in table~\ref{tab:wake}. In this section, the implications
of approximating the ULLT wake model/kernel are investigated by
comparing the normalized lift coefficient amplitude from all three
ULLTs against CFD for heaving
kinematics. Figure~\ref{fig:_kernel_C_L_comparison} shows the $|C_L| /
kh_0^*$ curves for the three aspect ratios of wing studied, across a
range of chord reduced frequencies.

All of the ULLTs predict the same low frequency $|C_L| / kh_0^*$ limit
- an expected property due to the interaction kernels of the C-ULLT
and the S-ULLT becoming equivalent to the P-ULLT kernel in the low
frequency limit. This limiting value reduces with aspect ratio of the
wing. For all \AR s in general, the prediction from C-ULLT is larger in
value than that from S-ULLT, which in turn is larger than the
prediction from P-ULLT. All methods show an initial negative slope in
the normalized $C_L$ amplitude until $k \approx 0.5$, and a positive
slope thereafter.

At high frequencies $k \approx > 1$, $|C_L| / kh_0^*$ has a linear
slope in all 3 ULLT. This is owing to the fact that the added mass
contribution to $|C_L| / kh_0^*$, which varies linearly with $k$ (see
equation~\ref{eq:2d_C_l_h}), is dominant at high frequencies. The curves
for the C-ULLT and the S-ULLT converge due to their identical high
frequency limiting behavior. With sufficiently high $k$, they will
converge with strip theory. The P-ULLT predicts a slightly lower
gradient. The interaction kernel
of the P-ULLT does not have a zero high frequency limit. Consequently,
even at high frequency there is an erroneous 3D downwash correction, 
which luckily results in a better agreement with the results from CFD in this regime.
This luck may be specific to the rectangular wing planforms
studied in this paper. For wing planforms where the assumptions of ULLT
are satisfied, this downwash would introduce error.

Looking specifically at the \AR 8 results presented in
figure~\ref{fig:_ar8_ULLT_comp}, initially the C-ULLT and the S-ULLT
give a good prediction of the CFD result and the P-ULLT under-predicts
the result. As frequency increases, the C-ULLT and the S-ULLT
over-predict $|C_L| / kh_0^*$ with the over-prediction increasing with
frequency. The P-ULLT provides the best match compared to the CFD
result out of the ULLTs at $k\geq1$.

As aspect ratios decrease, as shown in
figures~\ref{fig:_ar4_ULLT_comp} and~\ref{fig:_ar2_ULLT_comp} for \AR s
4 and 2 respectively, there are two major trends of note.

Firstly, the initial negative slope of the $|C_L| / kh_0^*$ curves
decreases, and there is a greater difference between the C-ULLT and
S-ULLT curves. The difference between the C-ULLT and the S-ULLT curves
represent the failure to correct for the changing wake $\gamma_{w_y}$
distribution with respect to span in the outer domain of the S-ULLT.
At lower aspect ratios, these corrections 
accounting for finite-wing effects become more important.

Secondly, the ULLTs appear to increasingly over-predict the $|C_L| /
kh_0^*$ low frequency limit with decreasing aspect ratio. This effect
is most prominent at aspect ratio 2, but explains why at aspect ratio
4 the S-ULLT unexpectedly appears superior to the C-ULLT. Lifting-line
theory is based on the assumption of high aspect ratio, so increasing
error with decreasing aspect ratio is expected.

Summarizing, all the ULLTs provide a better prediction of the CFD
results than the strip theory (2D Theodorsen) result. In regimes where
the basic ULLT assumptions of high \AR{} and low $k$ are satisfied, the
complete C-ULLT provides the best predictions. For lower \AR s at low
frequencies ($k \approx < 0.5$), the simplified S-ULLT provided better
predictions than C-ULLT. At high frequencies $k\approx>1$, for all aspect
ratios in general, the pseudosteady P-ULLT provides the best
results. This is because the 3D unsteady induced downwash in this
method doesn't tend to zero at high frequencies (unlike in the other
two ULLTs), which better reflects the results from Euler CFD.

\subsubsection{Comparison of wake topologies in the ULLT kernels}
\label{sec:_ULLT_kernel_result_comparison_wake_vort}

\begin{figure}
\centering
\begin{tabular}{ >{\centering\arraybackslash}m{0.1\textwidth}  >{\centering\arraybackslash}m{0.4\textwidth}  >{\centering\arraybackslash}m{0.4\textwidth}}
\toprule
& $\gamma_{w_y}/ Uh_0^*$ & $\gamma_{w_x}/ Uh_0^*$\\
\midrule
	CFD
	& \includegraphics[width=0.4 \textwidth]{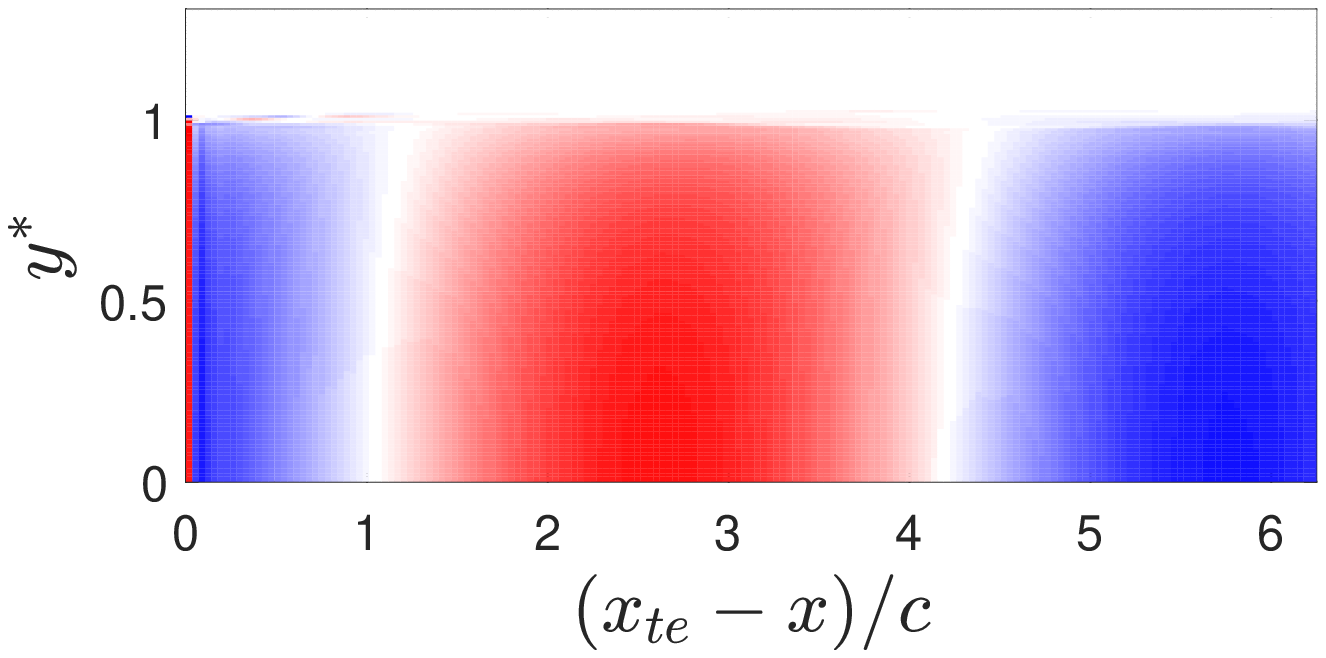} 
	& \includegraphics[width=0.4 \textwidth]{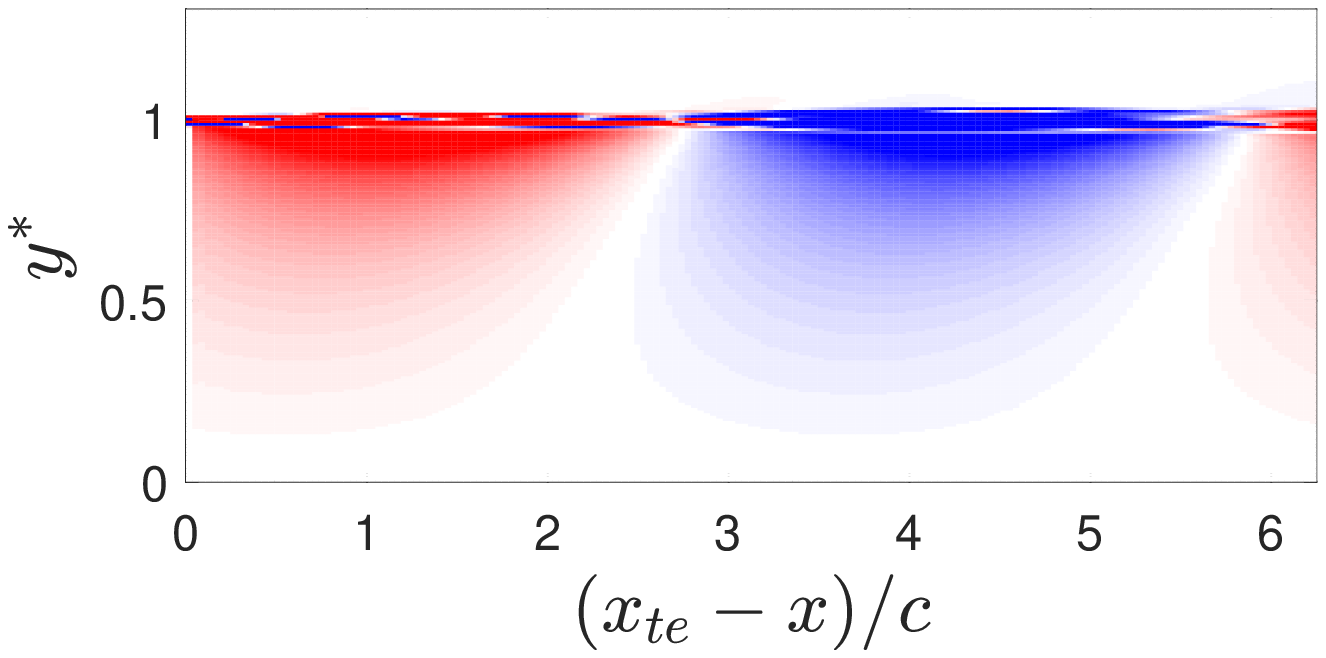} \\
	  
	C-ULLT
	&\includegraphics[width=0.4\textwidth]{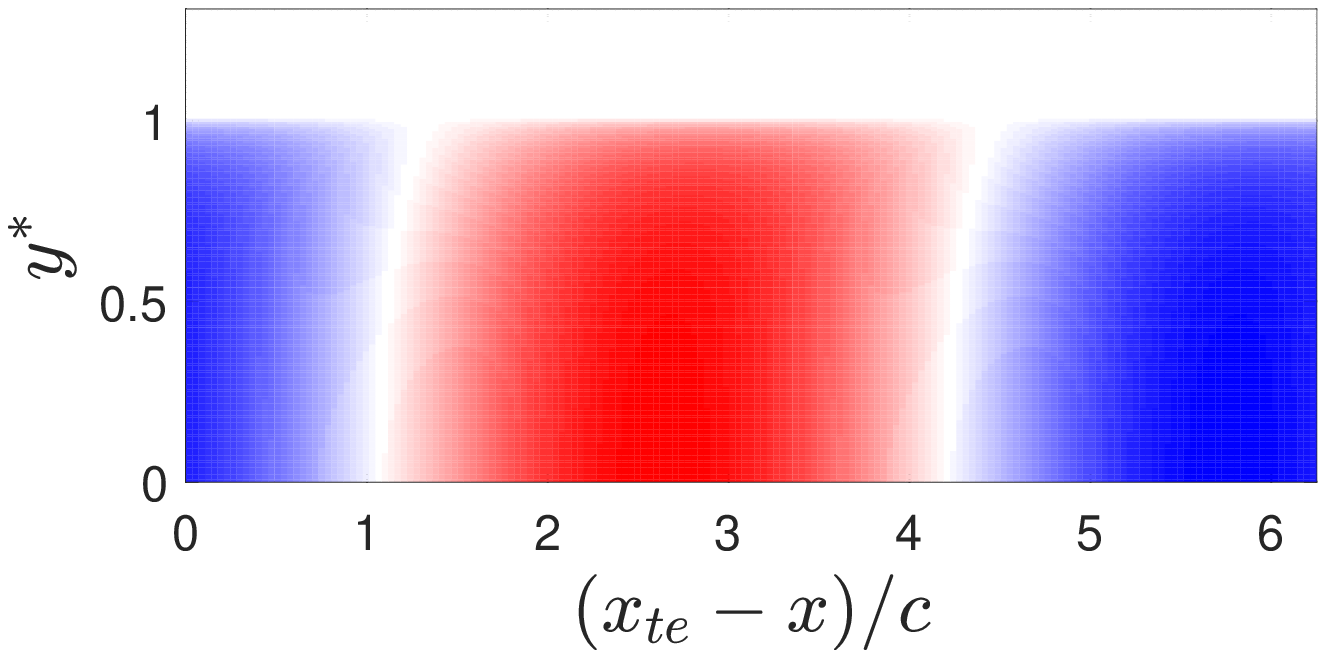}
	&\includegraphics[width=0.4\textwidth]{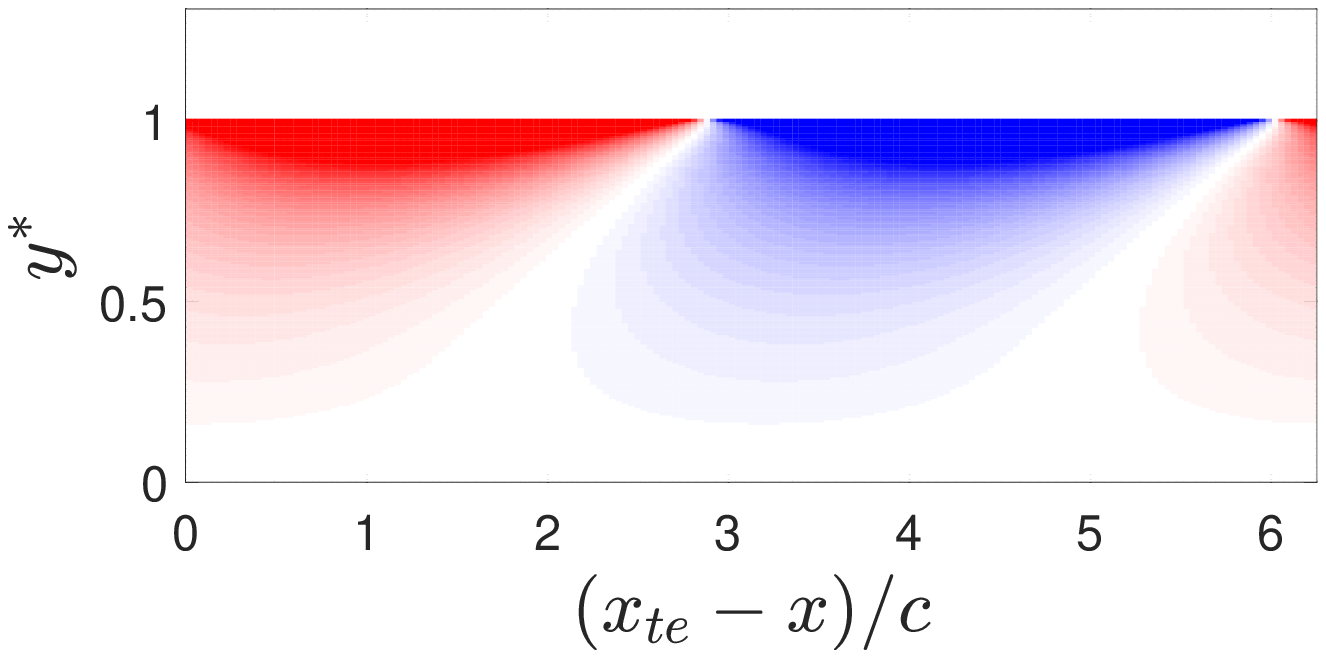}\\
	    	
	S-ULLT
	&\includegraphics[trim={-6cm -4cm -6cm -3cm},clip,width=0.4\textwidth]{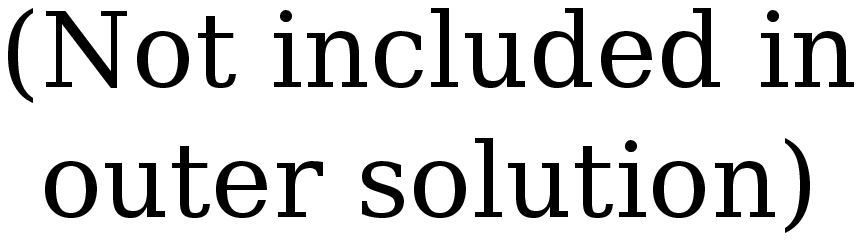}
	&\includegraphics[width=0.4\textwidth]{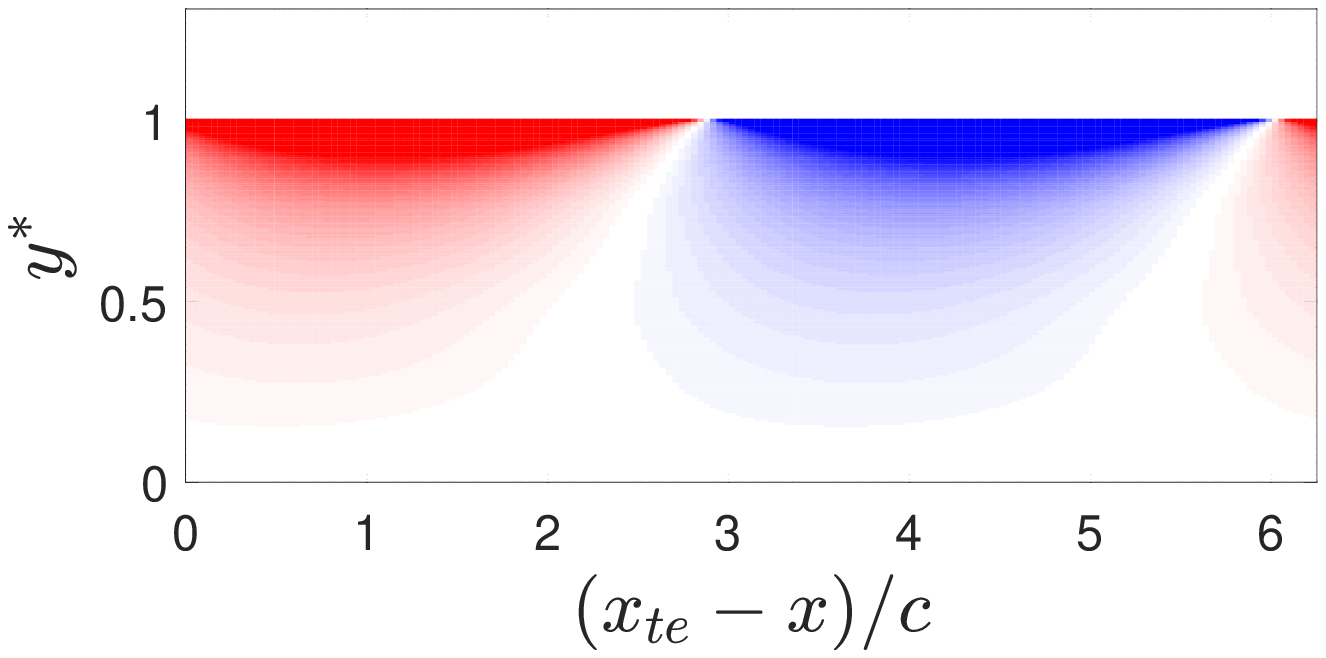}\\
	    	
          P-ULLT
	&\includegraphics[trim={-6cm -4cm -6cm -3cm},clip,width=0.4\textwidth]{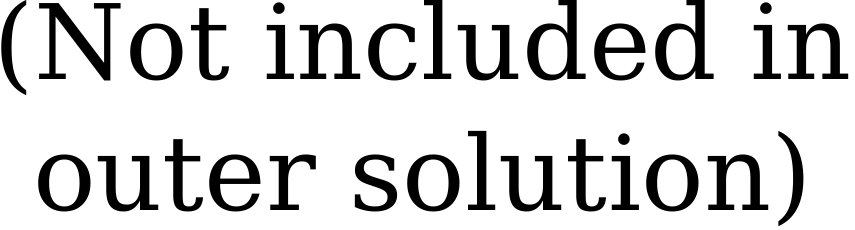}
	&\includegraphics[width=0.4\textwidth]{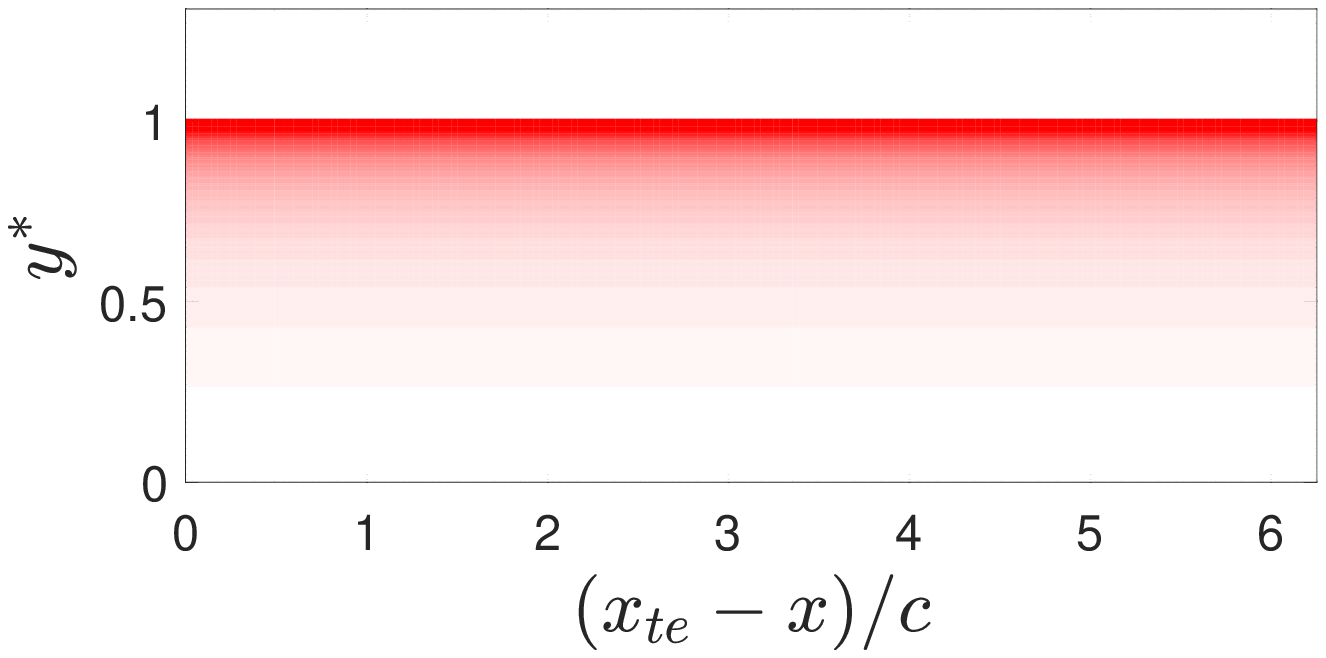}\\
\bottomrule
\end{tabular}
	    	\includegraphics[trim={0 0 0 8cm},clip,width=0.6\textwidth]{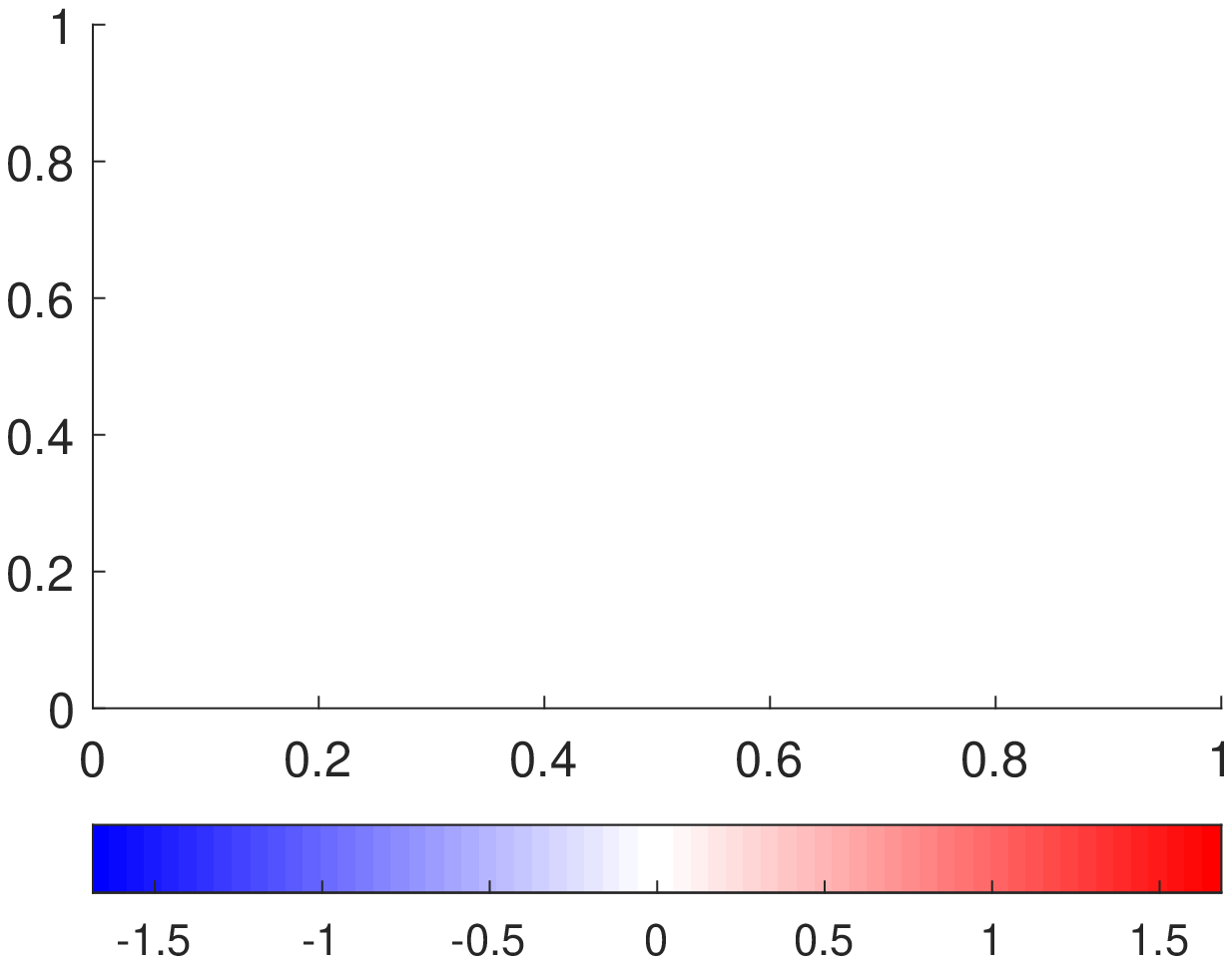}
		\caption{Comparison of wake vortex sheet strength for an aspect ratio 4 wing oscillating at $k=0.5$.
		The ULLT results show only the wake model assumed in the outer domain. Since the wing
		is shrunk to a line in the outer domain of the ULLTs, the $x$ coordinate of the trailing edge,
		$x_{te}$, is zero for ULLTs. Shown at $t \bmod 2\pi / \omega = 0$.}
		\label{fig: vorticity distributions k=0.5 aspect ratio 4}
\end{figure}

To better illustrate the differences between the wake approximations
in the three ULLTs, the vortex sheet strength in the wake from CFD was
compared against the vortex sheet strength assumed for the outer domain of
the ULLTs. This is shown in figure~\ref{fig: vorticity distributions
  k=0.5 aspect ratio 4} for the \AR 4 wing oscillating in heave with
$k=0.5$. 

To obtain the plotted data from the CFD cases, the vorticity was
integrated over a line in the $z$ direction at each point on the
$x$-$y$ plane. For ULLTs, the vorticity as assumed in the outer domain
is shown.

The C-ULLT predicts the CFD results well. Some error is introduced,
as would be expected, near to the wing tip. This error is most visible
in the $\gamma_{w_x}$ plots. In the C-ULLT, a singular distribution
of $\gamma_{w_x}$ is predicted due to the assumed remapped Fourier 
distribution of bound circulation. In the CFD, this streamwise wake
vorticity remains finite, instead spreading out. In the plots
of $\gamma_{w_y}$, the error near the wing tip in the C-ULLT
prediction is most visible when comparing the difference in phase
between the vorticity at the center of the span and at the tip. In the 
C-ULLT result, there is a larger phase difference than is observed 
in the CFD results.

The S-ULLT produces a similar $\gamma_{w_x}$ field to the
C-ULLT. There is a small phase difference between the C-ULLT
result and the S-ULLT result. At the wing tip, where the streamwise
vorticity has the greatest amplitude, the phase is very similar however.
The S-ULLT does not model $\gamma_{w_y}$ in the outer
domain. Whilst $\gamma_{w_y}$ is included in the inner domain, there
is no correction for its variation over the span. This applies to the
P-ULLT as well.

The P-ULLT also doesn't account for the sinusoidal variation of
$\gamma_{w_x}$ in the outer wake. However, it roughly matches the CFD
$\gamma_{w_x}$ distribution close to the wing (up to about 2 chord
lengths from the trailing edge). Since the vorticity closer to the
wing has a larger impact on the induced downwash, the P-ULLT obtains
reasonable solutions despite this assumption.

\subsubsection{Influence of ULLT kernel on spanwise lift distribution}
\label{sec:_lift_and_moment_dists}

An advantage of ULLT over strip theory is that it accounts for finite
wing effects (with bound circulation going to zero at the wingtips) when
computing force distributions over the wing. For aeroelastic analysis
(for which strip theory is often used), these load distributions are
of importance. This section compares the lift distributions obtained
using the three ULLTs against that obtained from Euler CFD.

The distribution of lift and changes with respect to oscillation
frequency. This occurs because of the growing importance of added
mass-effects with increasing oscillation frequency, and also because
of the dependence of the 3D interaction kernel $K$ on span reduced
frequency $\nu$. For the C-ULLT and the S-ULLT the 3D unsteady induced
downwash tends to zero as frequency increases.

To compare unsteady lifting-line theory to CFD, the lift distribution
was extracted from the CFD data at 8 equispaced time instants over a
single oscillation. A sine wave was then fitted to the data using a
least squares method to determine the amplitude.

The lift distributions will be compared for rectangular wings of
aspect ratios 8, 4 and 2 for heaving oscillations at three different
chord reduced frequencies. First, a low frequency $k=0.125$ where
C-ULLT provides the best agreement with CFD (so long as aspect ratio
is high) is studied. Next, a high frequency $k=1.5$ where added-mass
effects dominate and where P-ULLT was seen to make the best
predictions is studied. Finally, an intermediate frequency $k=0.5$
where the loads are influenced strongly by both circulatory and
added-mass effects is examined.

\vspace{10pt}\needspace{4\baselineskip}
\noindent\textbf{Low frequency behavior}

\begin{figure}
\centering
    \subfigure[Aspect ratio 8]{
	\label{fig: heave lift distribution comparison AR8 k=0.125}
    	\includegraphics[height=0.376\textwidth]{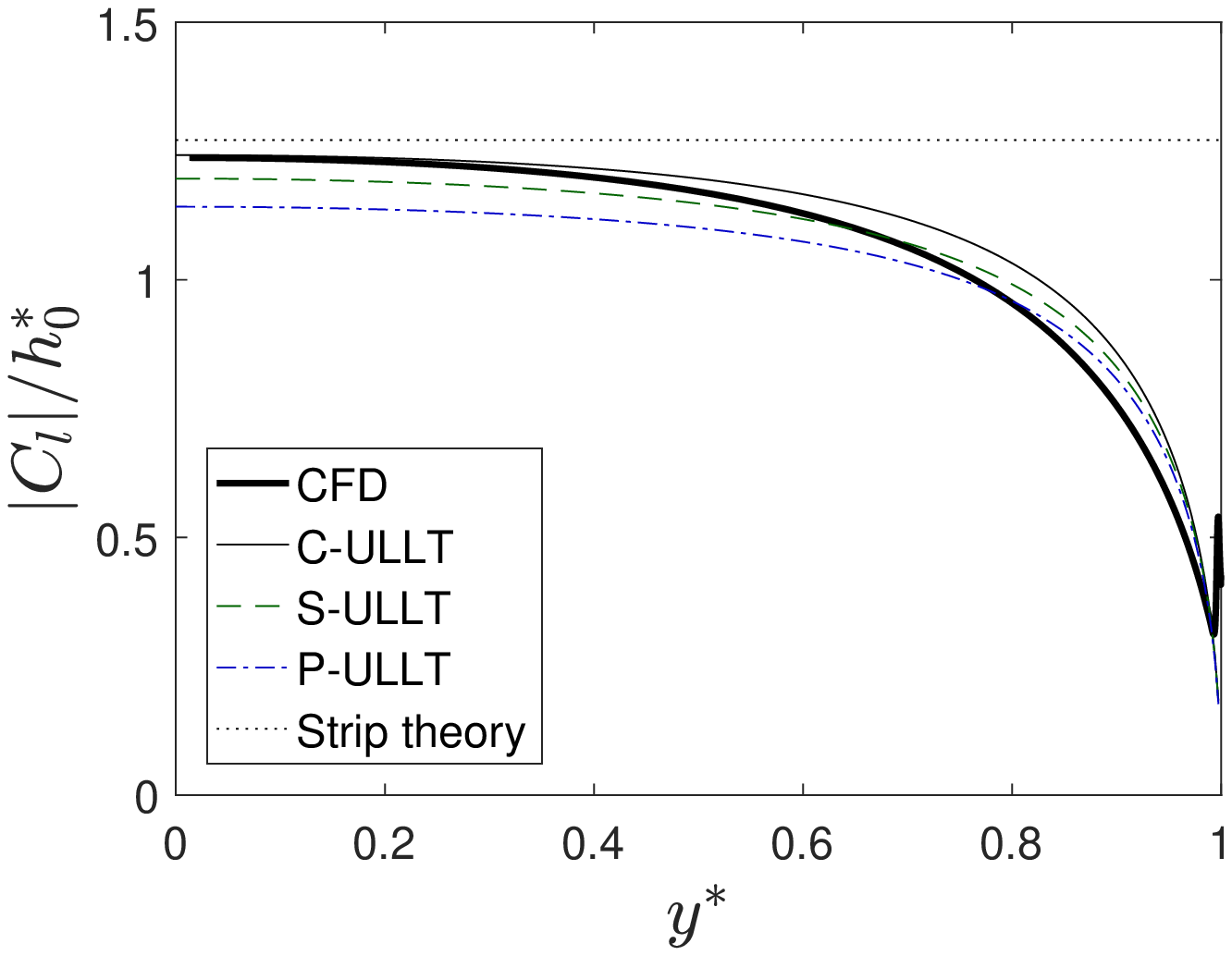}}
    \subfigure[Aspect ratio 4]{
	\label{fig: heave lift distribution comparison AR4 k=0.125}
    	\includegraphics[height=0.376\textwidth]{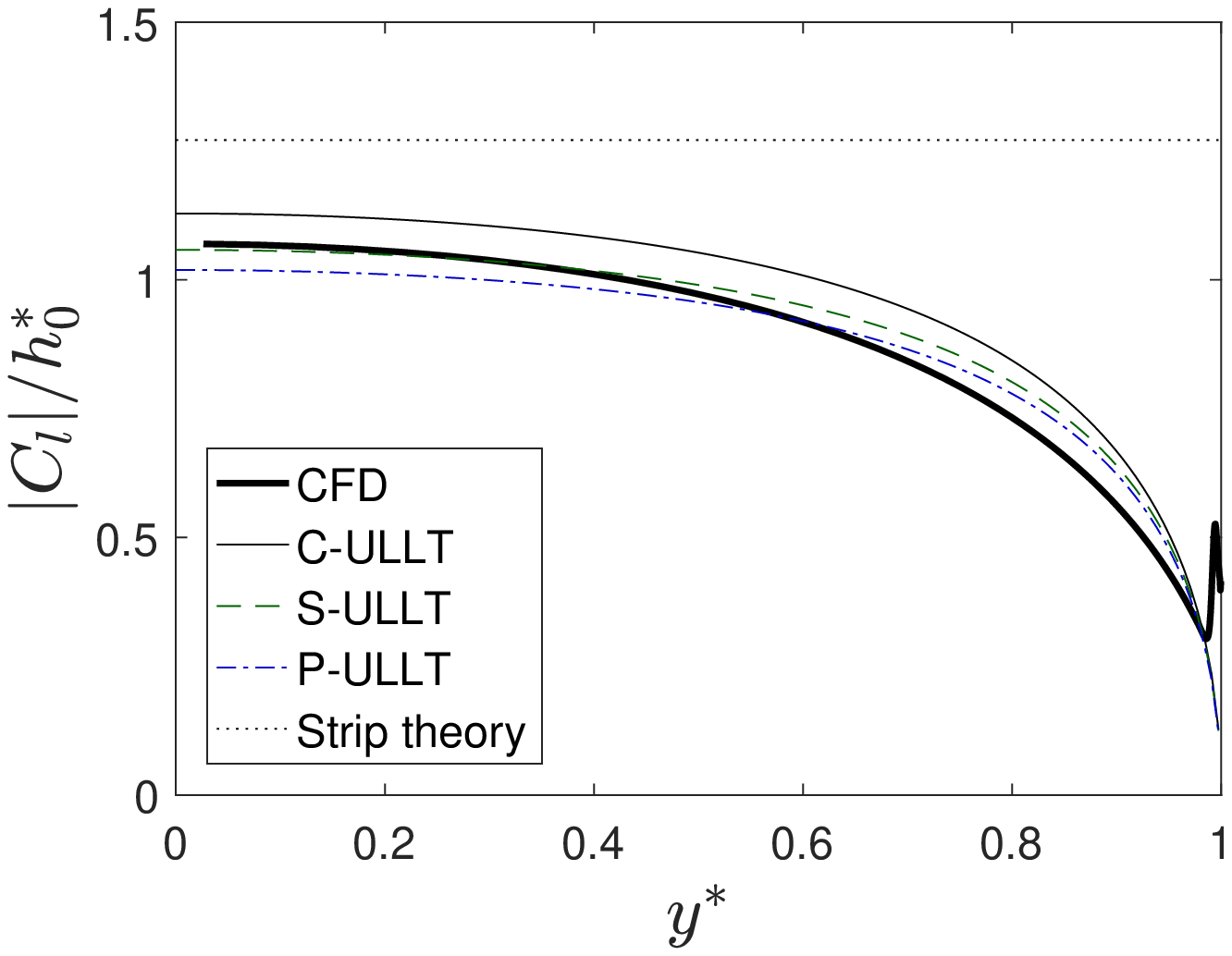}}
    
    \subfigure[Aspect ratio 2]{
	\label{fig: heave lift distribution comparison AR2 k=0.125}
    	\includegraphics[height=0.376\textwidth]{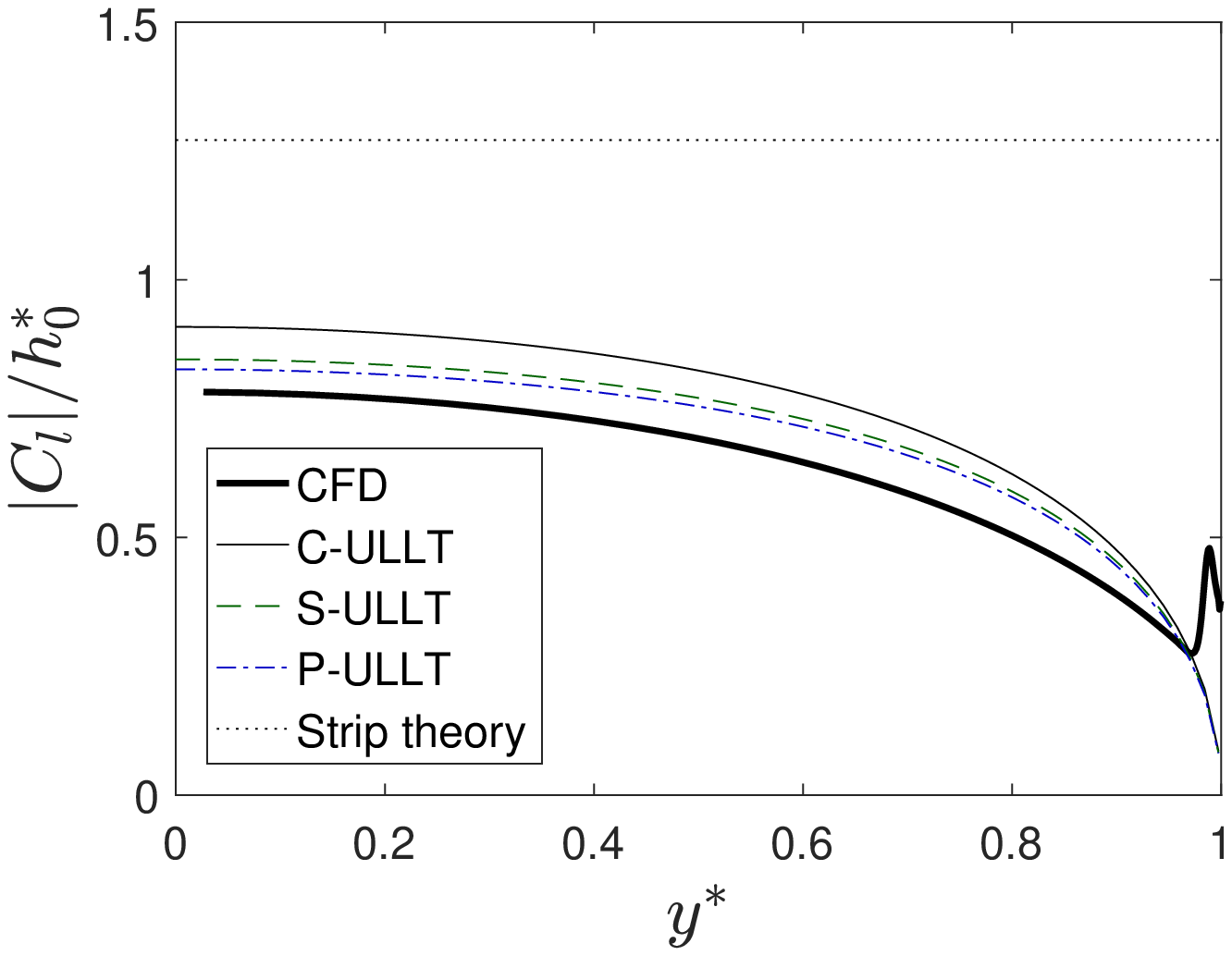}}
    
	\caption{Comparison of spanwise lift distribution from ULLTs,
          strip theory and CFD data with respect for wings oscillating
          in heave at $k=0.125$.}
	\label{fig: heave distributions comparison k=0.125}	
\end{figure}

The lift distribution amplitudes from ULLTs are compared against Euler
CFD and strip theory for the low frequency case ($k=0.125$) in
figure~\ref{fig: heave distributions comparison k=0.125}.

The CFD results show a mostly smooth distribution of lift coefficient,
with the maximum $|C_l|$ being found at the wing root ($y=0$). Near
the tip, $|C_l|$ starts to decrease ever more rapidly, excepting a
spike at the very wing tip. This is caused by separation at the sharp
edges of the wing tip.

The difference between the $|C_l|$ curves predicted
by the different ULLTs at $k=0.125$ is small due the fact that they
all tend to the P-ULLT kernel as frequency decreases. The results
obtained by the C-ULLT and S-ULLT are in particular more similar. The
C-ULLT predicts the highest $|C_l|$ followed by the S-ULLT and then
the P-ULLT.

At aspect ratio 8 where ULLT is expected to be most valid, the C-ULLT
matches the CFD result well in the center of the wing but has
increasing errors as the wing tip is approached. As aspect ratio
decreases, the C-ULLT over-predicts the $|C_l|$ at the center of the
wing. Consequently at aspect ratio 4 either the S-ULLT or the P-ULLT
provides the best prediction of the CFD result. At aspect ratio 2, all
of the ULLTs over-predict $|C_l|$. The P-ULLT, which always gives the
lowest $|C_l|$, is therefore is closest to the CFD result.  In all
cases, ULLT is superior to strip theory.

\vspace{10pt}
\noindent\textbf{High frequency behavior}

\begin{figure}
\centering
    \subfigure[Aspect ratio 8]{
	\label{fig: heave moment distribution comparison AR8 k=1.5}
    	\includegraphics[height=0.376\textwidth]{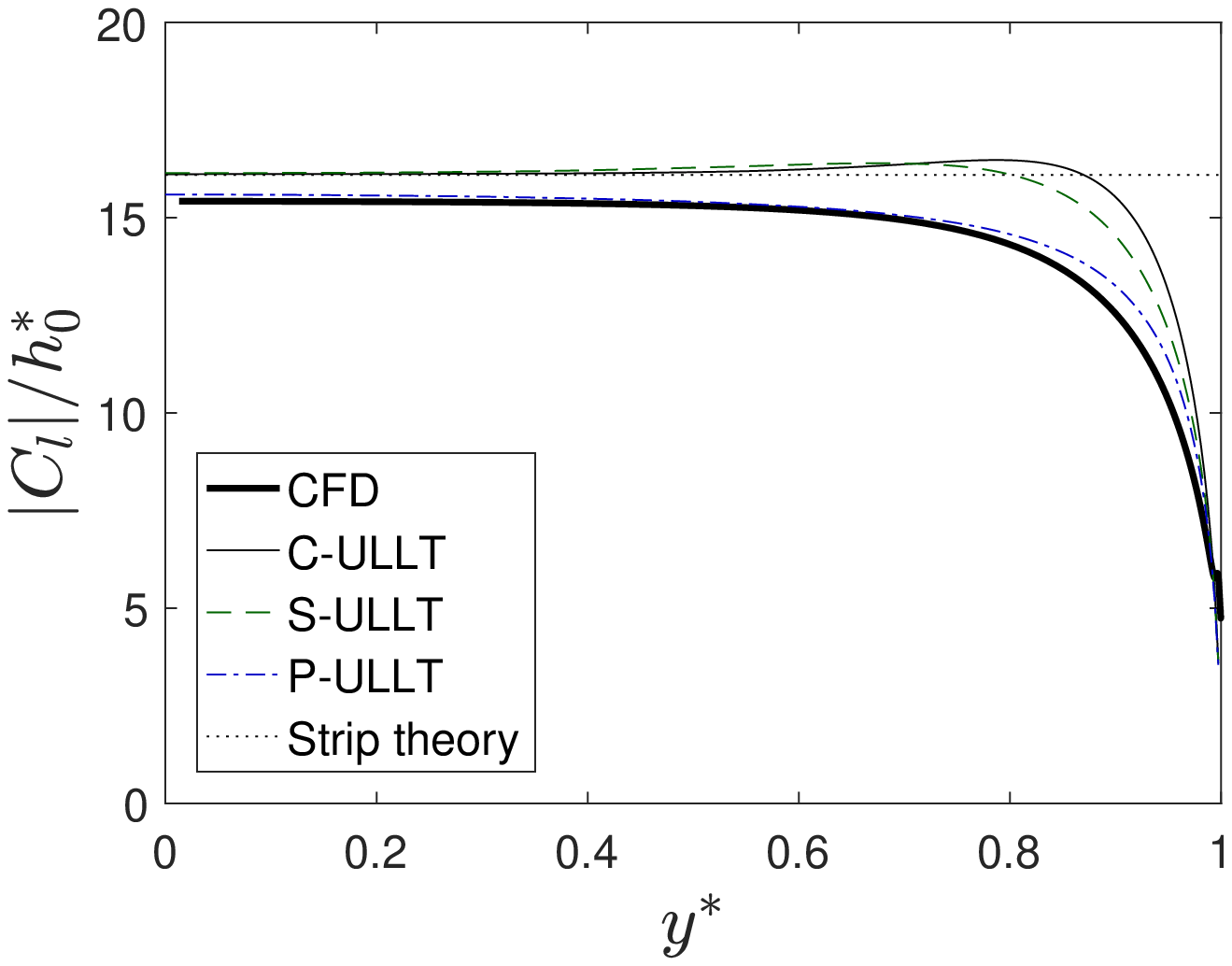}}
    \subfigure[Aspect ratio 4]{
	\label{fig: heave lift distribution comparison AR4 k=1.5}
    	\includegraphics[height=0.376\textwidth]{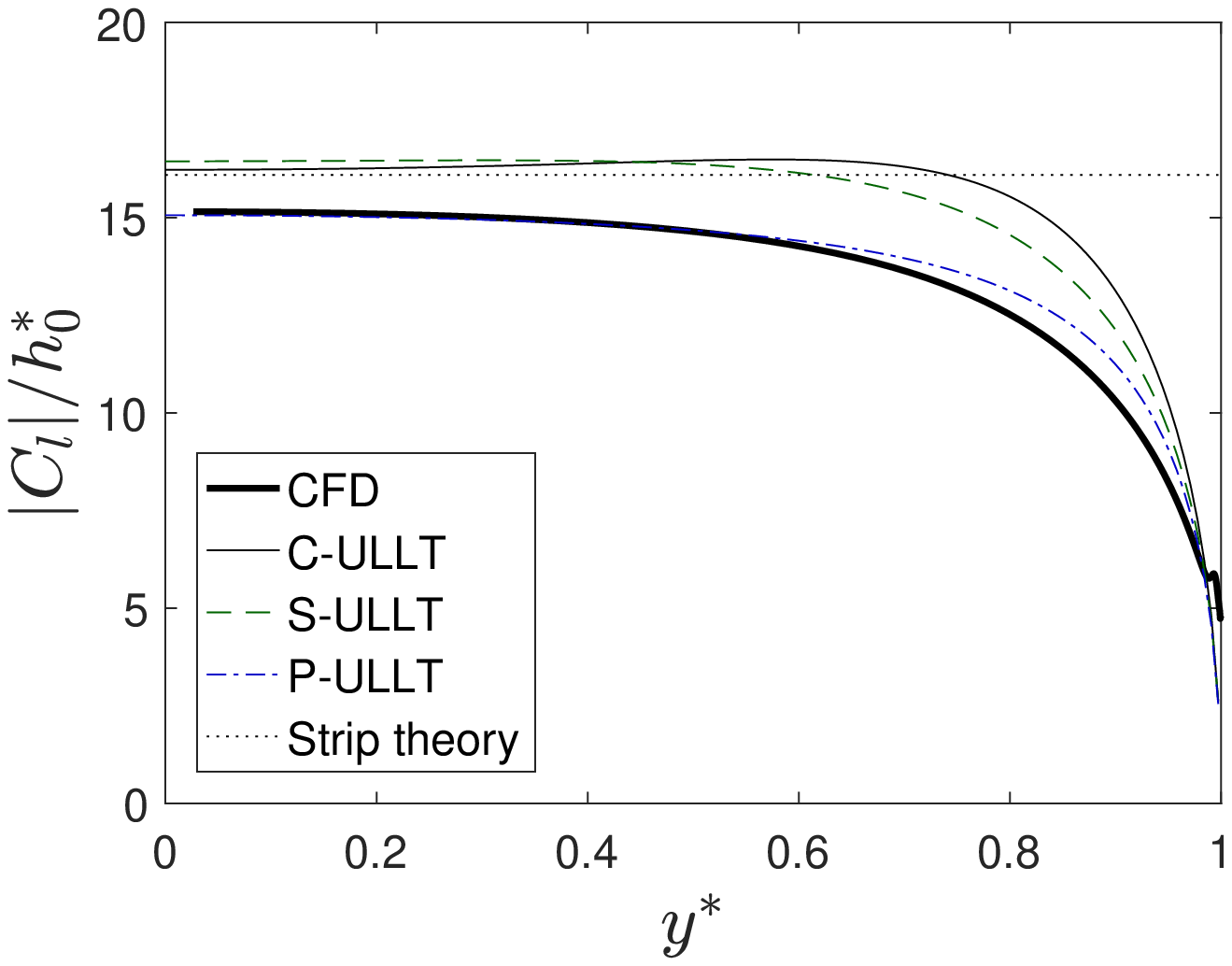}}

    \subfigure[Aspect ratio 2]{
	\label{fig: heave lift distribution comparison AR2 k=1.5}
    	\includegraphics[height=0.376\textwidth]{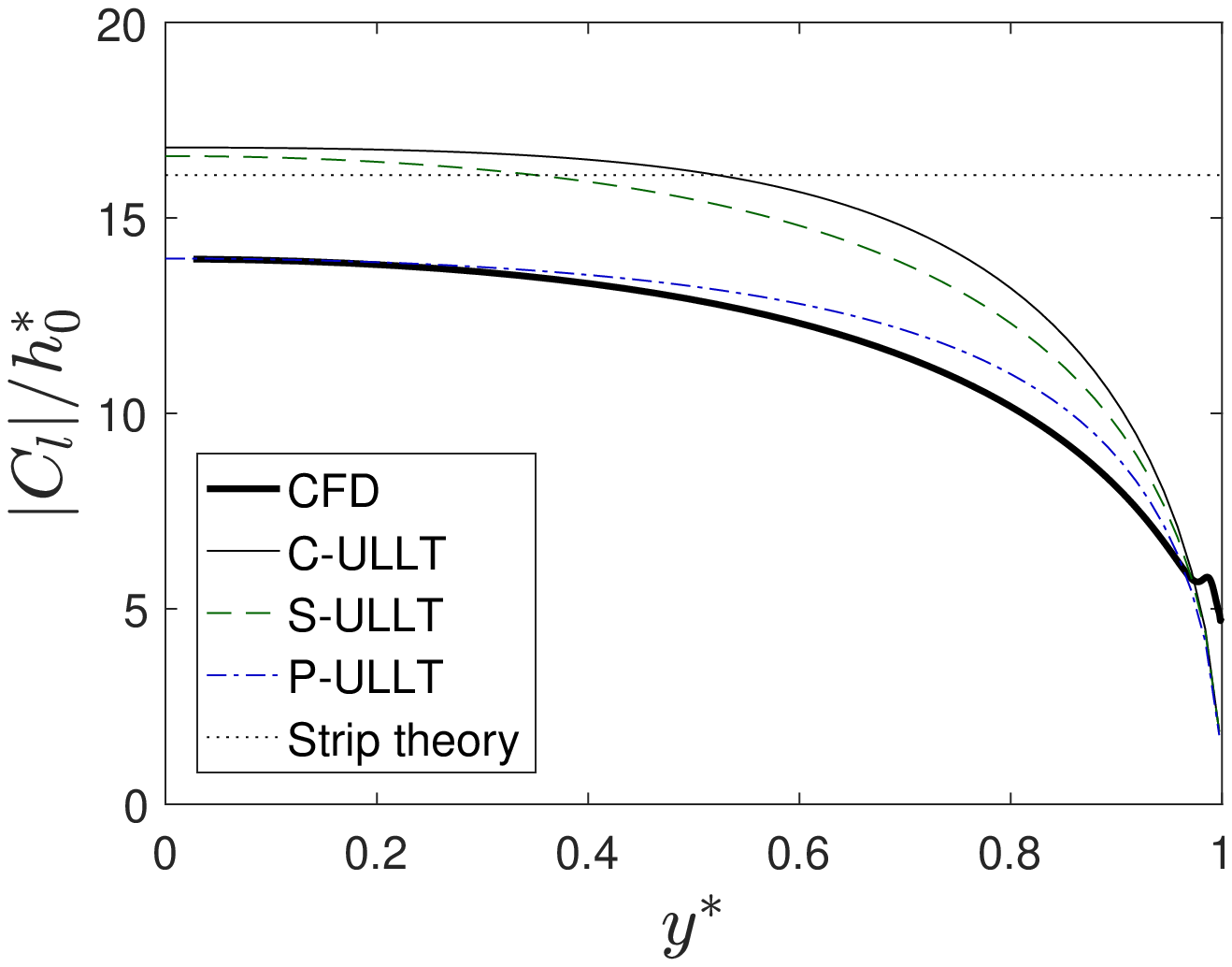}}

	\caption{Comparison of spanwise lift distribution from ULLTs,
          strip theory and CFD data with respect for wings oscillating
          in heave at $k=1.5$.}
	\label{fig: heave distributions comparison k=1.5}	
\end{figure}

Figure~\ref{fig: heave distributions comparison k=1.5} shows a
comparison of CFD and ULLT lift distribution predictions for the high
frequency case $k=1.5$.

In section~\ref{sec:_ULLT_kernel_result_comparison}, it has been noted
that the pseudosteady ULLT gives the best results at high frequencies, 
as the incorrectly modeled downwash inadvertently reduced the lift 
overestimate due to the lifting-line assumption being poor at rectangular 
wing tips. This is further confirmed in figure~\ref{fig: heave
distributions comparison k=1.5}; P-ULLT agrees best with CFD for all
three aspect ratios while C-ULLT and S-ULLT over-predict the lift
distribution. For \AR 8, the P-ULLT prediction matches with CFD over
most of the wing, until $y^* \approx 0.7$. As aspect ratio decreases,
the distance from the wingtip where errors are present increases, as
expected.

\newpage
\noindent\textbf{Intermediate frequency behavior}

\begin{figure}
\centering
    \subfigure[Aspect ratio 8, $k=0.5$]{
	\label{fig: heave moment distribution comparison AR8 k=0.5}
    	\includegraphics[height=0.376\textwidth]{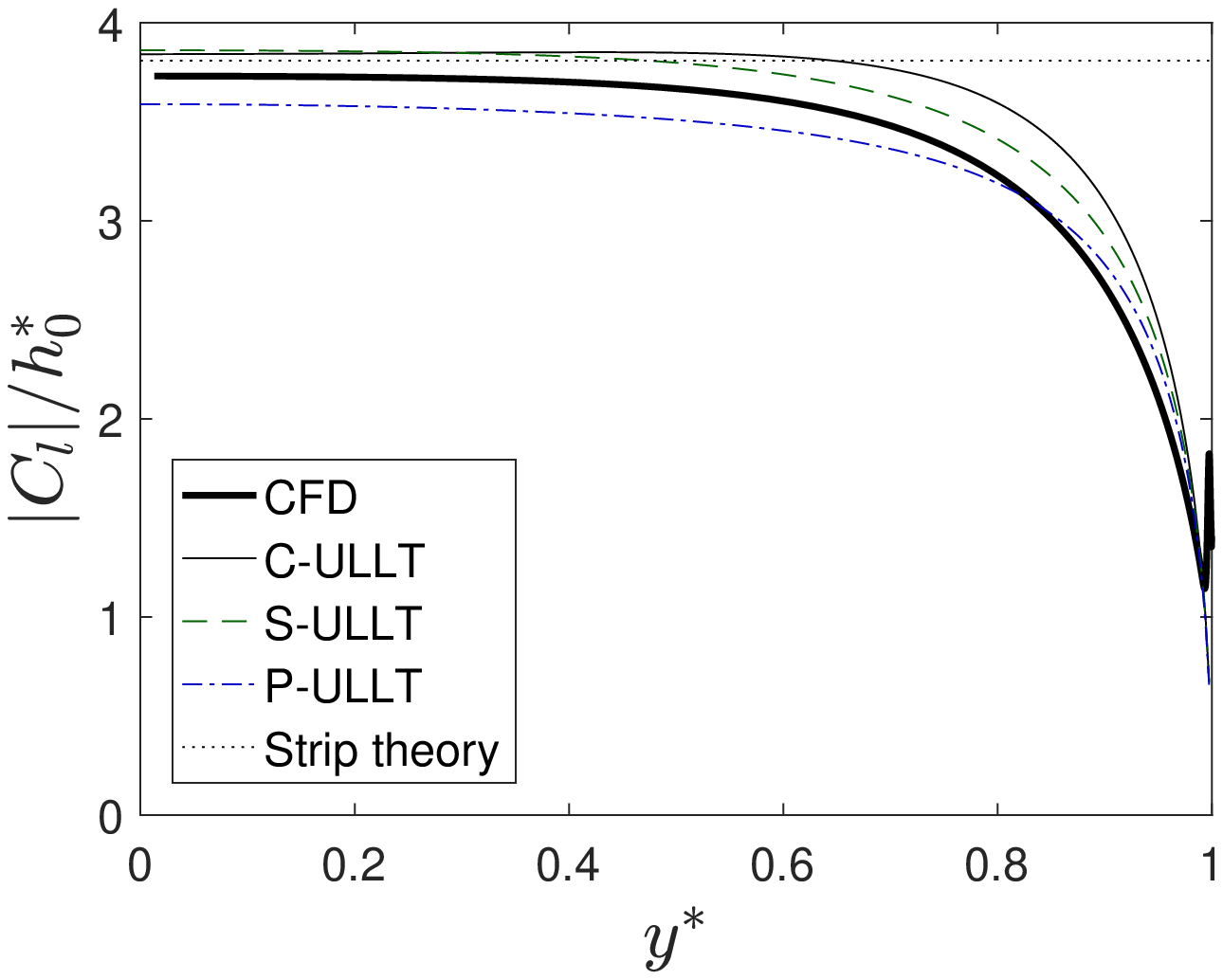}}
    \subfigure[Aspect ratio 4, $k=0.5$]{
	\label{fig: heave lift distribution comparison AR4 k=0.5}
    	\includegraphics[height=0.376\textwidth]{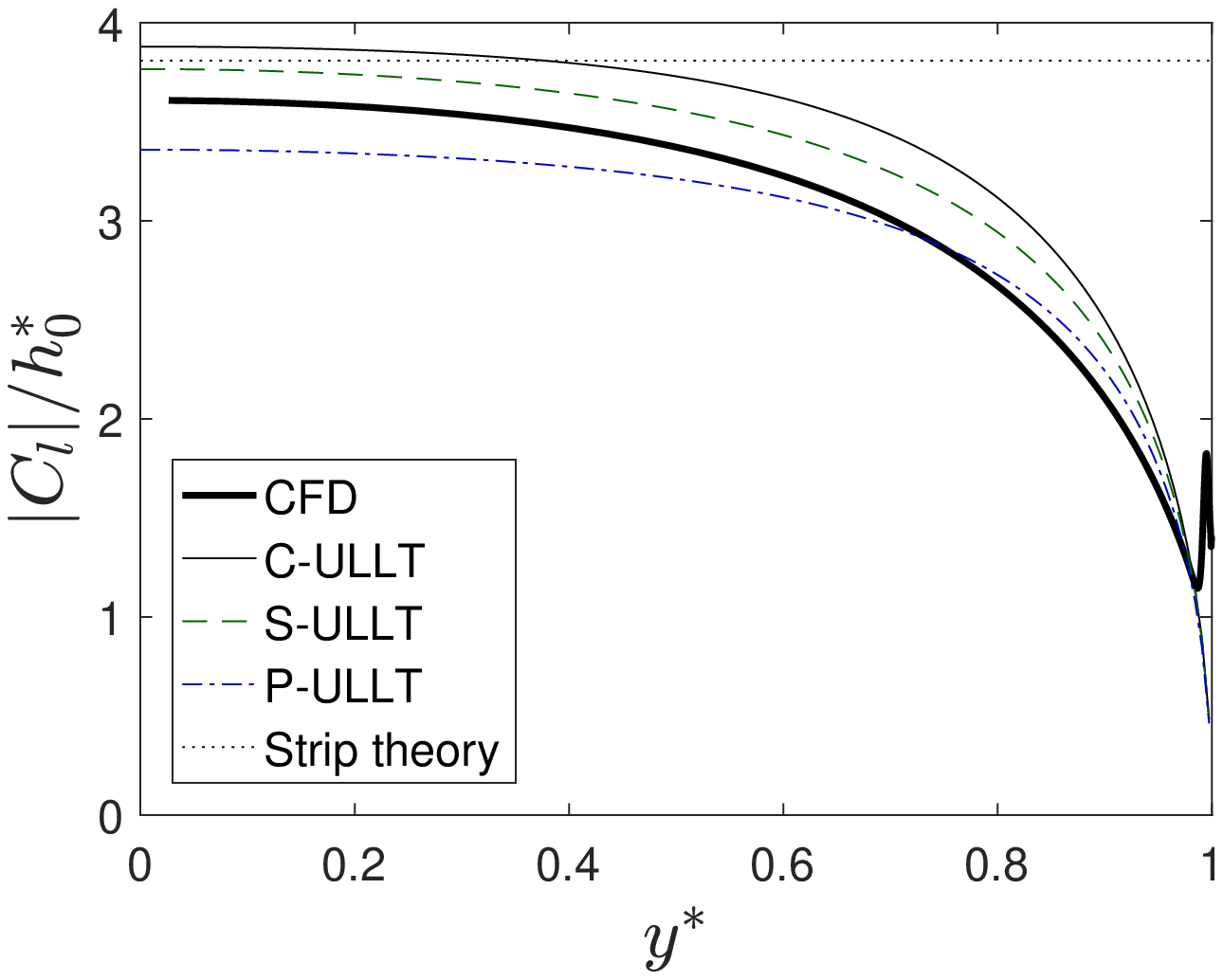}}
    
    \subfigure[Aspect ratio 2, $k=0.5$]{
	\label{fig: heave lift distribution comparison AR2 k=0.5}
    	\includegraphics[height=0.376\textwidth]{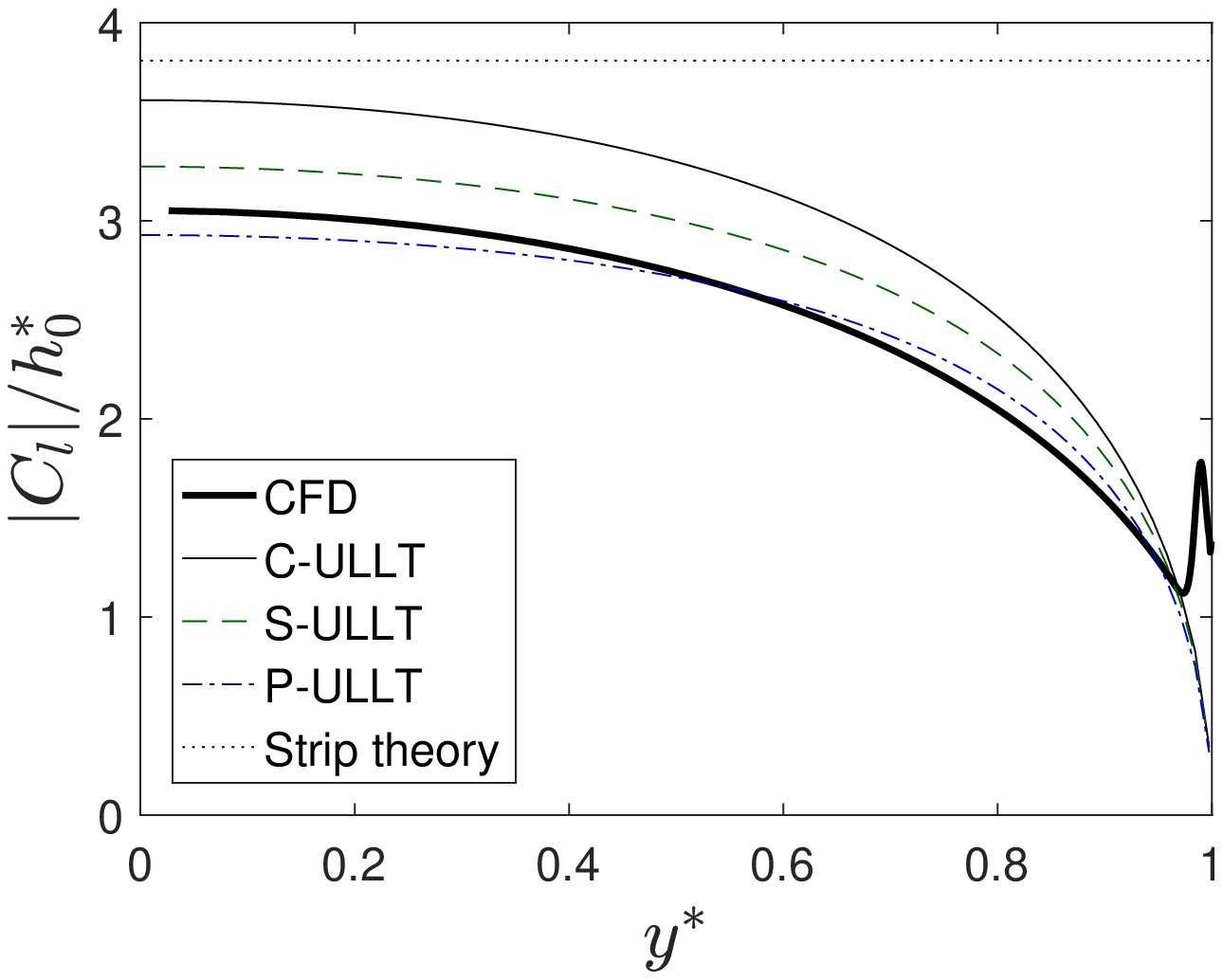}}
	\caption{Comparison of spanwise lift distribution from ULLTs,
          strip theory and CFD data with respect for wings oscillating
          in heave at $k=0.5$.}
	\label{fig: heave distributions comparison k=0.5}	
\end{figure}

In the intermediate frequency range, the choice of best ULLT model is
less clear. Figure~\ref{fig: heave distributions comparison k=0.5}
shows a comparison of CFD and ULLT lift distribution predictions at
$k=0.5$, where added-mass and circulatory effects are equally
important.

The C-ULLT lift distribution curves have the same shape as those from
CFD at all aspect ratios. However, there is an offset which increases
as aspect ratio decreases. The simplified S-ULLT which in general
predicts lower lift values than C-ULLT hence provides better
predictions for \AR s 8 and 4. P-ULLT which always predicts lower lift
values than both C-ULLT and S-ULLT, provides the best prediction for
AR2.

\subsubsection{Choice of ULLT kernel}
\label{sec:_wake_model_choice_advice}

The comparison of ULLT models against CFD in
section~\ref{sec:cfd_analysis} provides guidance on choosing the most
suitable ULLT depending on the problem parameters. In the regime where
the assumptions used in ULLT perturbation analysis are best satisfied,
i.e. at high aspect ratio and low chord reduced frequency, Sclavounos'
C-ULLT works best. On the other hand, at high chord reduced
frequencies, for all aspect ratios, the pseudosteady P-ULLT provides
the best predictions. P-ULLT also provides better predictions than the
other two ULLTs at low aspect ratio, across all frequencies. At higher
aspect ratios and at intermediate frequencies, the S-ULLT prediction
(which lies between the C-ULLT and P-ULLT predictions) provides the
best agreement with CFD. The superiority of the P-ULLT and S-ULLT
in comparison to the C-ULLT is likely due to the error from the 
rectangular wing tips. Strip theory was shown in all the results as
a reference to illustrate the importance of 3D effects (unsteady
induced downwash) in various regimes; ULLT universally provides better
predictions than strip theory.

All the ULLTs are easy to implement and have very low computational
cost in comparison with vortex lattice methods and CFD. C-ULLT and
S-ULLT have more complex kernels than P-ULLT. 
For numerical equivalents to these theories, a method similar
to the C-ULLT is most complex to implement, with Devinant \cite{Devinant1998}
showing how it is necessary to cancel the component of spanwise wake
vorticity present in both inner and outer domain. An method based on the S-ULLT alleviates
this complexity, whilst remaining more accurate than a pseudosteady method.

\subsection{Validation with experimental data}
\label{sec:exp_val}

In section~\ref{sec:cfd_analysis}, CFD simulations of the
incompressible Euler equations were used to validate unsteady
lifting-line theory and to compare the three kernels representing the
3D unsteady induced downwash. Here, ULLT is validated against against
experimental data (at finite Reynolds and Mach number) taken from NASA
technical report 4632\cite{Piziali1994} to confirm that the assumption
of inviscid flow does not invalidate its use for practical
applications.

NASA TR-4632\cite{Piziali1994} contains data for an aspect ratio
$10.1$ rectangular wing with a NACA0015 section undergoing pitch
oscillation about its quarter chord. The Reynolds number is $1.951$
million and the Mach number is $0.288$. A case with chord reduced
frequency $k=0.133$, average angle of attack $3.98\degree$, and pitch
amplitude $4.35\degree$ was selected. This data is presented in figure
90 of the report. In terms of physical quantities in the wind tunnel,
the wing had a span of 60.62 inches, chord of 12 inches, oscillation
frequency of 14.02Hz, and was subject to a free stream of 100.58
ms$^{-1}$.

The studies conducted in
sections~\ref{sec:_ULLT_kernel_result_comparison}
and~\ref{sec:_lift_and_moment_dists} indicate that for case described
above which falls in the high-aspect ratio, low-frequency regime, (i)
the C-ULLT provides the best prediction and (ii) the results from all
three ULLTs are nearly the same (see
figures~\ref{fig:_kernel_C_L_comparison} and~\ref{fig: heave
  distributions comparison k=0.125}). Hence, the experimental data is
compared against only the C-ULLT prediction below.

\begin{figure}
\centering \subfigure[$y^* = 0.250$]{
	\label{fig:_NASA_circles_y0p25}
    	\includegraphics[width=0.48\textwidth]{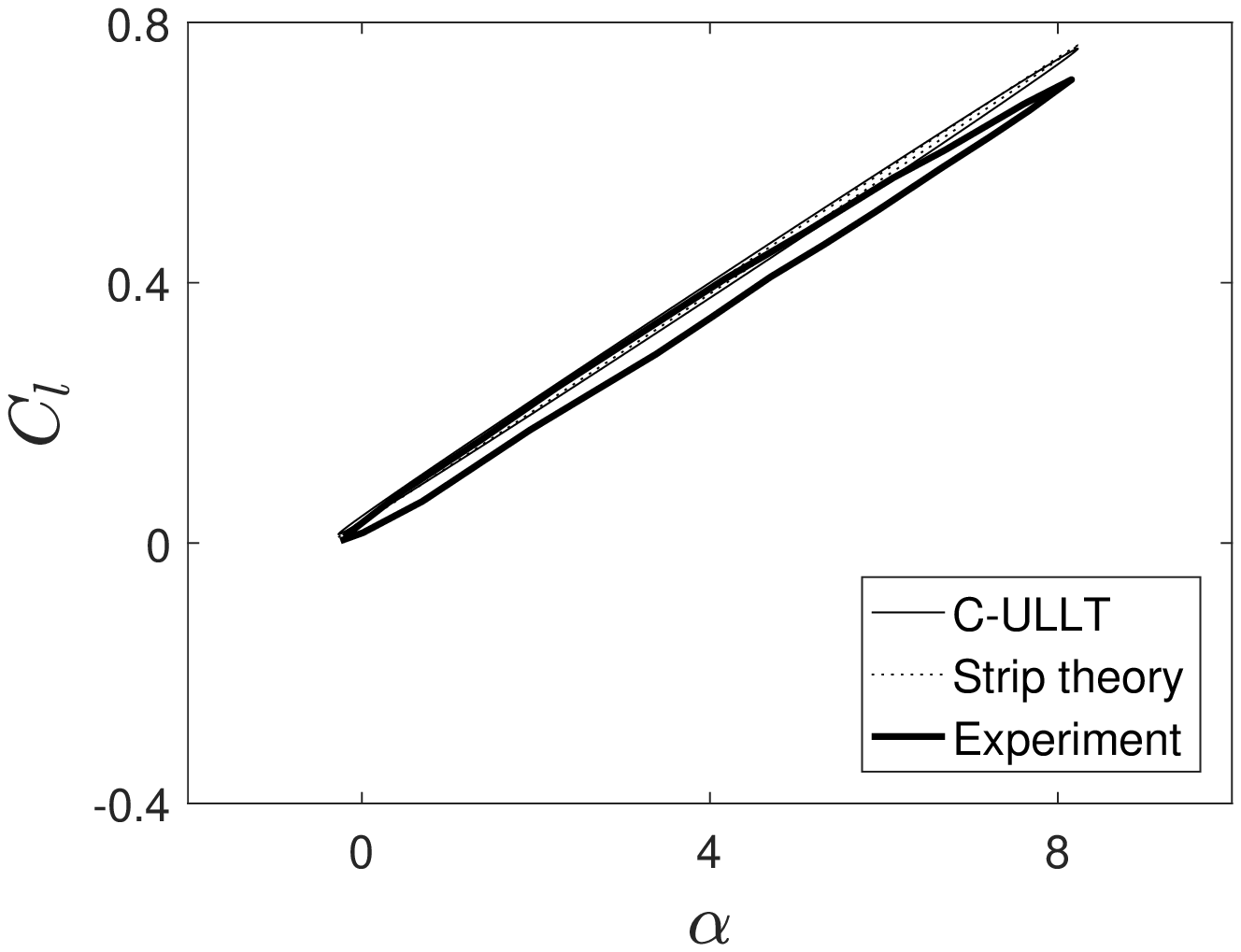}}    
    \subfigure[$y^* = 0.475$]{
	\label{fig:_NASA_circles_y0p475}
    	\includegraphics[width=0.48\textwidth]{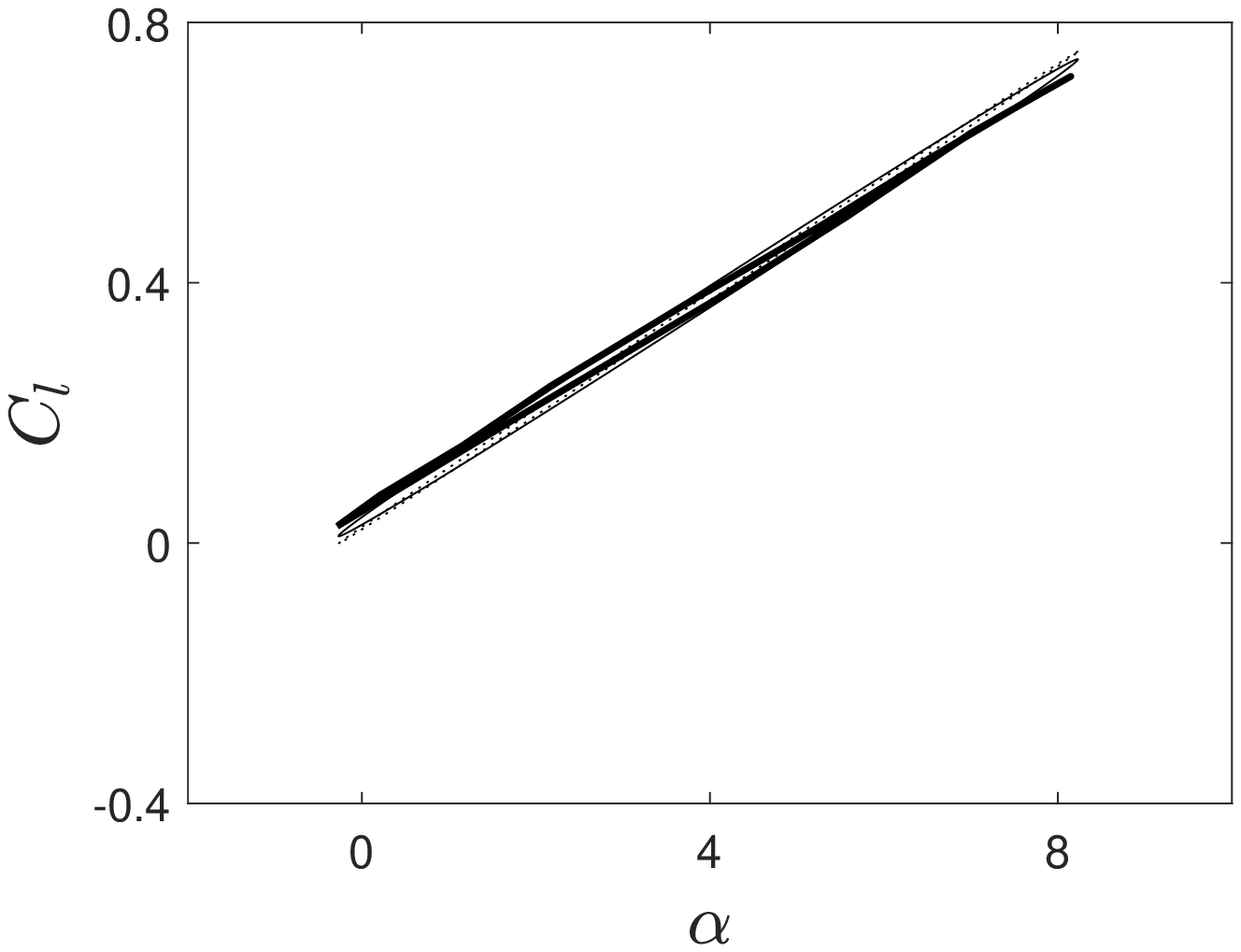}}
    	
    \subfigure[$y^* = 0.800$]{
	\label{fig:_NASA_circles_y0p800}
    	\includegraphics[width=0.48\textwidth]{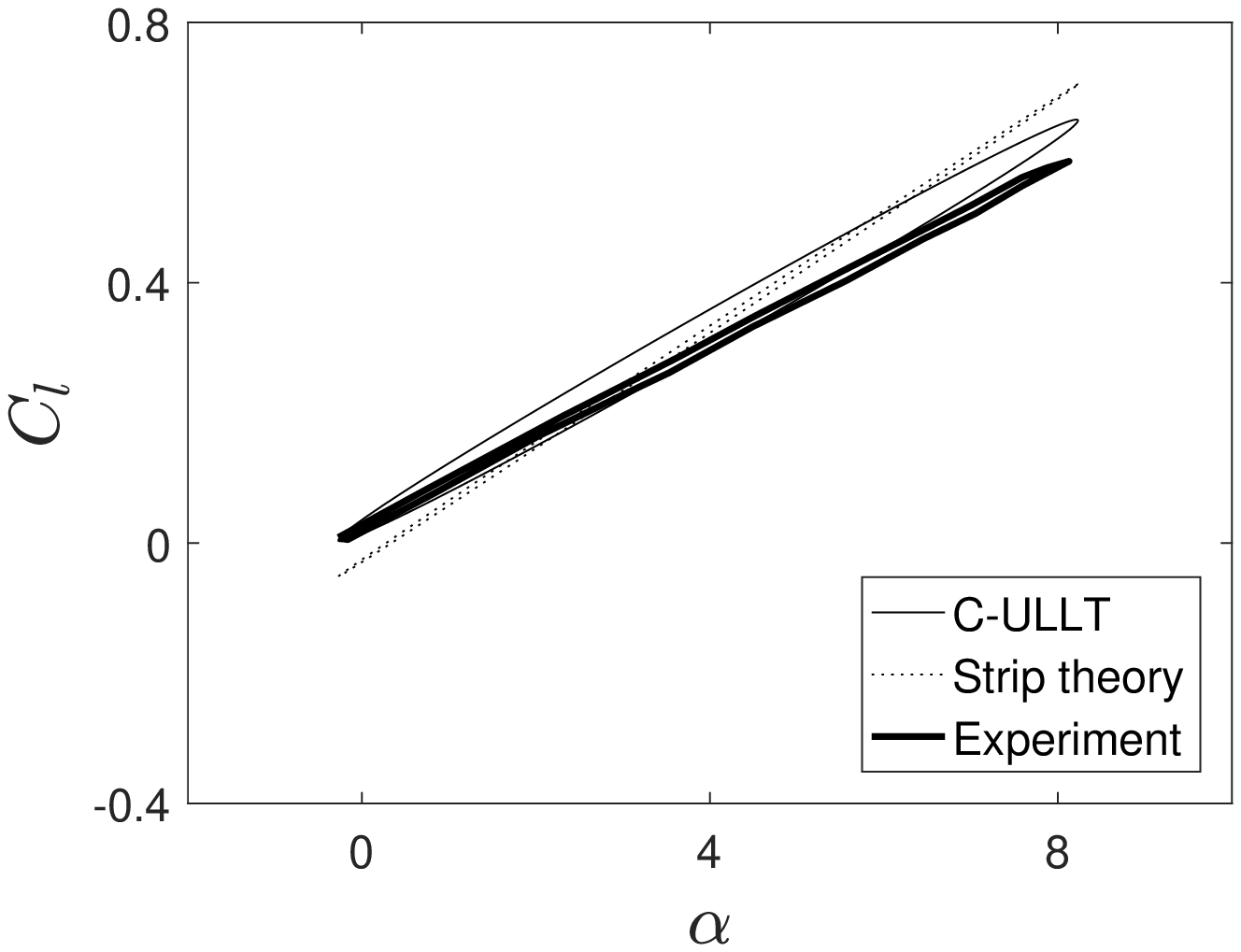}}		    
    \subfigure[$y^* = 0.966$]{
	\label{fig:_NASA_circles_y0p966}
    	\includegraphics[width=0.48\textwidth]{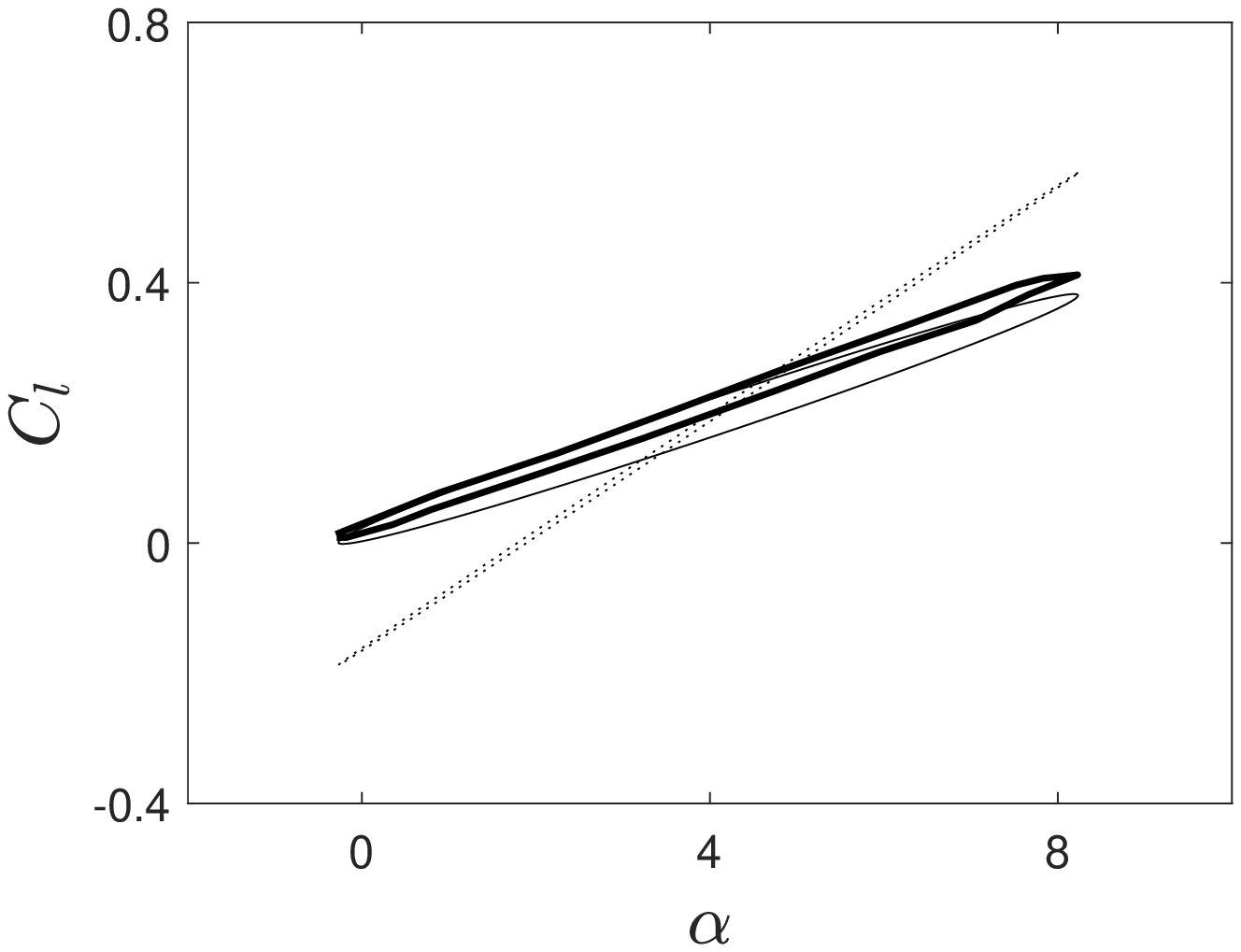}}

	\caption{A comparison of the lift coefficient predicted by ULLT/LLT, strip theory and experiment.}
	\label{fig:_NASA_TR4632_circles}	
\end{figure}

The ULLTs assume a zero average pitch angle. To model 
a non-zero average, the results of the ULLT are summed with the 
steady results of Prandtl’s lifting-line theory \cite{Prandtl1923}. This is consistent 
because both theories are linear. Variation of sectional lift coefficient ($C_l$)
with angle of attack, at different locations over the wing span, from
C-ULLT (with steady solution added) and experiment are compared in
figure~\ref{fig:_NASA_TR4632_circles}.

The experimental data has both the largest $C_l$ amplitude and the
highest average $C_l$ (compared to itself) nearest to the center of
the wing at $y^*=0.25$ as shown in
figure~\ref{fig:_NASA_circles_y0p25}. There is a small phase
difference between the angle of attack and lift coefficient, resulting
in an elliptic curve.  At $y^*=0.475$, shown in
figure~\ref{fig:_NASA_circles_y0p475}, the amplitude of $C_l$ remains
similar, although the less elliptical curves suggests that the loads
are now more closely in phase to the kinematics.

Closer to the tip at $y^*=0.800$, the amplitude of the $C_l$ obtained
by the experiment is reduced. Figure~\ref{fig:_NASA_circles_y0p966}
shows the result very to the wing tip at $y^*=0.966$. Both $C_l$
average and amplitude are further reduced in comparison to the inboard
result. The curve obtained in noticeably non-smooth.  
These results agree with the distribution of $|C_l|$ obtained using
CFD for lower aspect ratios. Namely, that the $|C_l|$ amplitude
decreases near the wing tips.

Near the center of the wing at $y^*=0.25$ and $y^*=0.475$
(figures~\ref{fig:_NASA_circles_y0p25} and
\ref{fig:_NASA_circles_y0p475}), the match between theory (ULLT) and
experiment is very good. Strip theory, shown for comparison, also
provides very good agreement with CFD at these spanwise
locations. This is in line with expectations for very high aspect
ratio wings ($10$ in this case), where the behavior near the root
would be nearly 2D.

Moving towards the tip, the improvement offered by ULLT over strip
theory is clear. The ULLT/LLT slightly over-predicts both the mean and
amplitude of $C_l$ at $y^*=0.8$, shown in
figure~\ref{fig:_NASA_circles_y0p800}, and also suggests a larger
phase difference between the kinematics and loads than experiment. At
$y^*=0.986$, the ULLT/LLT under-predicts $C_l$. However, at both these
spanwise locations near the wing tip, ULLT provides a much better
prediction than strip theory.

%% file: 090-summary.tex
\section{Conclusions}
\label{sec: conclusions}

In this paper, three unsteady lifting-line theories (ULLTs) have been obtained from a common framework:
a `complete' ULLT contains corrections for both components of vorticity in the outer domain, a novel simplified 
version that considers only the streamwise component, and a pseudosteady, Prandtl-like ULLT.

These ULLTs were systematically compared both against each other and  against Euler CFD results
for various aspect ratio rectangular wings at various oscillation frequencies. 
The ULLTs were capable of capturing trends in wing lift coefficient with respect to both
oscillation frequency and aspect ratio, always providing better results than 2D strip theory.
They were also able to capture the changes in lift distribution across the wing with respect to 
frequency, despite being less accurate close to the wing tips. For aeroelastic analysis,
this provides advantages over assumed lift-distribution models unable to account for changes
in oscillation frequency.

Comparing between the three ULLTs, it was found that in regimes where ULLT assumptions
are best satisfied (high aspect ratio and low reduced frequency), the complete ULLT is
most accurate. At high frequencies of motion, the Prandtl-like pseudosteady ULLT 
provided the best prediction, although this may be luck specific to the
rectangular wings studied.

By modifying the wake model, the computational
cost could be reduced at the expense of reduced accuracy. The novel ULLT considering only the 
streamwise component of the vorticity in the outer domain was found to provide an excellent
reduction in computational cost with minimal loss in accuracy. The pseudosteady ULLT
reduced computational cost further still, but provided worse accuracy at low frequencies. 

ULLT was also compared against experimental data, confirming that it is not only applicable
to Euler/inviscid problems, but also to practical high Reynolds number flows.

The results obtained in this paper are important because they show how ULLT
provides a low computational cost model capable of accounting for the interacting
3D aerodynamic effects of aspect ratio and oscillation frequency. 
The simplified wake models provide not only
a means by which accuracy can be traded for reductions in computational cost,
but also guidance on the construction of numerical and time-domain ULLTs. In
particular, the streamwise vorticity ULLT demonstrates how an entire component
of outer solution wake vorticity can be neglected (along
with its self-canceling singularities), with only a small cost to the accuracy of 
the solution.

%% file: 095-ack.tex
\section{Acknowledgments}

The authors gratefully acknowledge the support of the UK Engineering and Physical Sciences Research Council
(EPSRC) through a DTA scholarship and grant EP/R008035. The Cirrus UK National Tier-2 HPC service at EPCC
(http://www.cirrus.ac.uk) and the ARCHIE-WeSt 
High Performance Computer (www.archie-west.ac.uk) based at the University of Strathclyde
were used in CFD simulations. We'd also like to thank the reviewers of this paper for their insightful feedback.

\section*{Data Availability}

The datasets generated during and/or analysed during the current study are available from the corresponding author on reasonable request.

%% file: 000-springer.bbl
\begin{thebibliography}{10}
\providecommand{\url}[1]{{#1}}
\providecommand{\urlprefix}{URL }
\expandafter\ifx\csname urlstyle\endcsname\relax
  \providecommand{\doi}[1]{DOI~\discretionary{}{}{}#1}\else
  \providecommand{\doi}{DOI~\discretionary{}{}{}\begingroup
  \urlstyle{rm}\Url}\fi

\bibitem{Ahmadi1985}
Ahmadi, A.R., Widnall, S.E.: Unsteady lifting-line theory as a singular
  perturbation problem.
\newblock Journal of Fluid Mechanics \textbf{153}, 59 (1985).
\newblock \doi{10.1017/s0022112085001148}

\bibitem{andreu2019influence}
Andreu~Angulo, I., Ansell, P.J.: Influence of aspect ratio on dynamic stall of
  a finite wing.
\newblock AIAA journal pp. 2722--2733 (2019)

\bibitem{beals2015lift}
Beals, N., Jones, A.R.: Lift production by a passively flexible rotating wing.
\newblock AIAA Journal \textbf{53}(10), 2995--3005 (2015)

\bibitem{biler2019experimental}
Biler, H., Badrya, C., Jones, A.R.: Experimental and computational
  investigation of transverse gust encounters.
\newblock AIAA Journal \textbf{57}(11), 4608--4622 (2019)

\bibitem{Bird2019}
Bird, H.J.A., Otomo, S., Ramesh, K., Viola, I.M.: A geometrically non-linear
  time-domain unsteady lifting-line theory.
\newblock In: {AIAA} Scitech 2019 Forum. American Institute of Aeronautics and
  Astronautics (2019).
\newblock \doi{10.2514/6.2019-1377}

\bibitem{bird2018theoretical}
Bird, H.J.A., Ramesh, K.: Theoretical and computational studies of a
  rectangular finite wing oscillating in pitch and heave.
\newblock Proceedings of the 6th. European Conference on Computational
  Mechanics (ECCM 6) and the 7th. European Conference on Computational Fluid
  Dynamics (ECFD 7) pp. 3944--3955 (2018)

\bibitem{Boutet2018}
Boutet, J., Dimitriadis, G.: Unsteady lifting line theory using the wagner
  function for the aerodynamic and aeroelastic modeling of 3d wings.
\newblock Aerospace \textbf{5}(3), 92 (2018).
\newblock \doi{10.3390/aerospace5030092}

\bibitem{calderon2014absence}
Calderon, D., Cleaver, D., Gursul, I., Wang, Z.: On the absence of asymmetric
  wakes for periodically plunging finite wings.
\newblock Physics of Fluids \textbf{26}(7), 349--376 (2014)

\bibitem{calderon2013lift}
Calderon, D., Wang, Z., Gursul, I.: Lift-enhancing vortex flows generated by
  plunging rectangular wings with small amplitude.
\newblock AIAA journal \textbf{51}(12), 2953--2964 (2013)

\bibitem{calderon2013volumetric}
Calderon, D.E., Wang, Z., Gursul, I., Visbal, M.: Volumetric measurements and
  simulations of the vortex structures generated by low aspect ratio plunging
  wings.
\newblock Physics of Fluids \textbf{25}(6), 067,102 (2013)

\bibitem{Caprace2020}
Caprace, D.G., Winckelmans, G., Chatelain, P.: An immersed lifting and dragging
  line model for the vortex particle-mesh method.
\newblock Theoretical and Computational Fluid Dynamics  (2020)

\bibitem{carr1988progress}
Carr, L.: Progress in analysis and prediction of dynamic stall.
\newblock Journal of Aircraft \textbf{25}(1), 6--17 (1988)

\bibitem{carr2}
Carr, L.W., Platzer, M.F., Chandrasekhara, M.S., Ekaterinaris, J.: Experimental
  and computational studies of dynamic stall.
\newblock In: T.~Cebeci (ed.) Numerical and Physical Aspects of Aerodynamic
  Flows IV, pp. 239--256. Springer Berlin Heidelberg (1990)

\bibitem{carr2013finite}
Carr, Z.R., Chen, C., Ringuette, M.J.: Finite-span rotating wings:
  three-dimensional vortex formation and variations with aspect ratio.
\newblock Experiments in fluids \textbf{54}(2), 1--26 (2013)

\bibitem{carr2015aspect}
Carr, Z.R., DeVoria, A.C., Ringuette, M.J.: Aspect-ratio effects on rotating
  wings: circulation and forces.
\newblock Journal of Fluid Mechanics \textbf{767}, 497--525 (2015)

\bibitem{cheng1976lifting}
Cheng, H.K.: On lifting-line theory in unsteady aerodynamics.
\newblock Tech. Rep. 133, University of Southern California Los Angeles, Dept.
  of Aerospace Engineering (1976)

\bibitem{corkery2018development}
Corkery, S., Babinsky, H., Harvey, J.: On the development and early
  observations from a towing tank-based transverse wing--gust encounter test
  rig.
\newblock Experiments in Fluids \textbf{59}(9), 135 (2018)

\bibitem{Darakananda2018}
Darakananda, D., Eldredge, J.D.: A versatile taxonomy of low-dimensional vortex
  models for unsteady aerodynamics.
\newblock Journal of Fluid Mechanics \textbf{858}, 917--948 (2018).
\newblock \doi{10.1017/jfm.2018.792}

\bibitem{Devinant1998}
Devinant, P.: An approach for unsteady lifting-line time-marching numerical
  computation.
\newblock International Journal for Numerical Methods in Fluids \textbf{26}(2),
  177--197 (1998).
\newblock
  \doi{10.1002/(sici)1097-0363(19980130)26:2<177::aid-fld633>3.3.co;2-g}

\bibitem{devoria2017mechanism}
DeVoria, A.C., Mohseni, K.: On the mechanism of high-incidence lift generation
  for steadily translating low-aspect-ratio wings.
\newblock Journal of Fluid Mechanics \textbf{813}, 110--126 (2017)

\bibitem{ekaterinaris1998computational}
Ekaterinaris, J.A., Platzer, M.F.: Computational prediction of airfoil dynamic
  stall.
\newblock Progress in aerospace sciences \textbf{33}(11), 759--846 (1998)

\bibitem{Eldredge2015}
Eldredge, J., Darakananda, D.: Reduced-order two- and three-dimensional vortex
  modeling of unsteady separated flows.
\newblock In: 53rd {AIAA} Aerospace Sciences Meeting. American Institute of
  Aeronautics and Astronautics (2015).
\newblock \doi{10.2514/6.2015-1749}

\bibitem{fishman2017structure}
Fishman, G., Wolfinger, M., Rockwell, D.: The structure of a trailing vortex
  from a perturbed wing.
\newblock Journal of Fluid Mechanics \textbf{824}, 701--721 (2017)

\bibitem{Gallay2015}
Gallay, S., Laurendeau, E.: Nonlinear generalized lifting-line coupling
  algorithms for pre/poststall flows.
\newblock {AIAA} Journal \textbf{53}(7), 1784--1792 (2015).
\newblock \doi{10.2514/1.j053530}

\bibitem{garrick1938}
Garrick, I.E.: On some reciprocal relations in the theory of nonstationary
  flows.
\newblock Tech. rep., NACA (1938)

\bibitem{green2011unsteady}
Green, M.A., Rowley, C.W., Smits, A.J.: The unsteady three-dimensional wake
  produced by a trapezoidal pitching panel.
\newblock Journal of Fluid Mechanics \textbf{685}, 117--145 (2011)

\bibitem{Guermond1990}
Guermond, J.L.: A generalized lifting-line theory for curved and swept wings.
\newblock Journal of Fluid Mechanics \textbf{211}(-1), 497 (1990).
\newblock \doi{10.1017/s0022112090001665}

\bibitem{Guermond1991}
Guermond, J.L., Sellier, A.: A unified unsteady lifting-line theory.
\newblock Journal of Fluid Mechanics \textbf{229}, 427 (1991).
\newblock \doi{10.1017/s0022112091003099}

\bibitem{Hansen2006}
Hansen, M., S{\o}rensen, J., Voutsinas, S., S{\o}rensen, N., Madsen, H.: State
  of the art in wind turbine aerodynamics and aeroelasticity.
\newblock Progress in Aerospace Sciences \textbf{42}(4), 285--330 (2006).
\newblock \doi{10.1016/j.paerosci.2006.10.002}

\bibitem{hirato2019vortex}
Hirato, Y., Shen, M., Gopalarathnam, A., Edwards, J.R.: Vortex-sheet
  representation of leading-edge vortex shedding from finite wings.
\newblock Journal of Aircraft pp. 1--15 (2019)

\bibitem{Holten1976}
Holten, T.V.: Some notes on unsteady lifting-line theory.
\newblock Journal of Fluid Mechanics \textbf{77}(03), 561 (1976).
\newblock \doi{10.1017/s0022112076002255}

\bibitem{hord2016leading}
Hord, K., Lian, Y.: Leading edge vortex circulation development on finite
  aspect ratio pitch-up wings.
\newblock AIAA Journal pp. 2755--2767 (2016)

\bibitem{James1975}
James, E.C.: Lifting-line theory for an unsteady wing as a singular
  perturbation problem.
\newblock Journal of Fluid Mechanics \textbf{70}(04), 753 (1975).
\newblock \doi{10.1017/s0022112075002339}

\bibitem{jantzen2014vortex}
Jantzen, R.T., Taira, K., Granlund, K., Ol, M.V.: Vortex dynamics around
  pitching plates.
\newblock Physics of Fluids \textbf{26}(5), 053,606 (2014)

\bibitem{Jones1939}
Jones, R.T.: The unsteady lift of a finite wing.
\newblock Tech. rep., National advisory committee for aeronautics (1939)

\bibitem{Katz2001}
Katz, J., Plotkin, A.: Low Speed Aerodynamics, 2 edn.
\newblock Cambridge University Press (2001)

\bibitem{Leishman2006}
Leishman, G.J.: Principles of Helicopter Aerodynamics.
\newblock Cambridge University Press (2006)

\bibitem{mancini2015unsteady}
Mancini, P., Manar, F., Granlund, K., Ol, M.V., Jones, A.R.: Unsteady
  aerodynamic characteristics of a translating rigid wing at low {R}eynolds
  number.
\newblock Physics of Fluids \textbf{27}(12), 123,102 (2015)

\bibitem{mccroskey}
McCroskey, W.J.: {The Phenomenon of Dynamic Stall}.
\newblock NASA TM 81264 (1981)

\bibitem{mcgowan2011investigations}
McGowan, G.Z., Granlund, K., Ol, M.V., Gopalarathnam, A., Edwards, J.R.:
  {Investigations of lift-based pitch-plunge equivalence for airfoils at low
  Reynolds numbers}.
\newblock AIAA Journal \textbf{49}(7), 1511--1524 (2011)

\bibitem{medina2016leading}
Medina, A., Jones, A.R.: Leading-edge vortex burst on a low-aspect-ratio
  rotating flat plate.
\newblock Physical Review Fluids \textbf{1}(4), 044,501 (2016)

\bibitem{moored2018unsteady}
Moored, K.W.: Unsteady three-dimensional boundary element method for
  self-propelled bio-inspired locomotion.
\newblock Computers \& Fluids \textbf{167}, 324--340 (2018)

\bibitem{mulleners2017flow}
Mulleners, K., Mancini, P., Jones, A.R.: Flow development on a flat-plate wing
  subjected to a streamwise acceleration.
\newblock AIAA Journal \textbf{55}(6), 2118--2122 (2017)

\bibitem{Murua2012a}
Murua, J., Palacios, R., Graham, J.M.R.: Applications of the unsteady
  vortex-lattice method in aircraft aeroelasticity and flight dynamics.
\newblock Progress in Aerospace Sciences \textbf{55}, 46--72 (2012).
\newblock \doi{10.1016/j.paerosci.2012.06.001}

\bibitem{ol2009shallow}
Ol, M.V., Bernal, L., Kang, C.K., Shyy, W.: {Shallow and deep dynamic stall for
  flapping low Reynolds number airfoils}.
\newblock Experiments in Fluids \textbf{46}(5), 883--901 (2009)

\bibitem{Olver2010}
Olver, F.W., Lozier, D.W., Boisvert, R.F., Clark, C.W.: {NIST} Handbook of
  Mathematical Functions Paperback and {CD-ROM}.
\newblock Cambridge University Press (2010)

\bibitem{ozen2012three}
Ozen, C.A., Rockwell, D.: Three-dimensional vortex structure on a rotating
  wing.
\newblock Journal of Fluid Mechanics \textbf{707}, 541--550 (2012)

\bibitem{perrotta2017unsteady}
Perrotta, G., Jones, A.R.: Unsteady forcing on a flat-plate wing in large
  transverse gusts.
\newblock Experiments in Fluids \textbf{58}(8), 101 (2017)

\bibitem{Piziali1994}
Piziali, R.A.: 2-d and 3-d oscillating wing aerodynamics for a range of angles
  of attack including stall.
\newblock Tech. rep., NASA (1994).
\newblock \urlprefix\url{https://ntrs.nasa.gov/search.jsp?R=19950012704}

\bibitem{Prandtl1923}
Prandtl, L.: Applications of modern hydrodynamics to aeronautics.
\newblock Tech. rep., NACA (1923).
\newblock Rep. 116

\bibitem{ramesh2020leading}
Ramesh, K.: On the leading-edge suction and stagnation point location in
  unsteady flows past thin aerofoils.
\newblock Journal of Fluid Mechanics \textbf{886}(A13) (2020)

\bibitem{kiran_journal1}
Ramesh, K., Gopalarathnam, A., Edwards, J.R., Ol, M.V., Granlund, K.: An
  unsteady airfoil theory applied to pitching motions validated against
  experiment and computation.
\newblock Theoretical and Computational Fluid Dynamics \textbf{27}(6), 843--864
  (2013)

\bibitem{ramesh2014discrete}
Ramesh, K., Gopalarathnam, A., Granlund, K., Ol, M.V., Edwards, J.R.:
  Discrete-vortex method with novel shedding criterion for unsteady airfoil
  flows with intermittent leading-edge vortex shedding.
\newblock Journal of Fluid Mechanics \textbf{751}, 500--538 (2014)

\bibitem{Ramesh2017}
Ramesh, K., Monteiro, T.P., Silvestre, F.J., Bernardo, A., aes Neto, G.,
  de~Souza Siqueria~Versiani, T., da~Silva, R.G.A.: Experimental and numerical
  investigation of post-flutter limit cycle oscillations on a cantilevered flat
  plate.
\newblock In: International Forum on Aeroelasticity and Structural Dynamics
  2017 (2017).
\newblock \urlprefix\url{http://eprints.gla.ac.uk/154722/}

\bibitem{Roesler2018}
Roesler, B.T., Epps, B.P.: Discretization requirements for vortex lattice
  methods to match unsteady aerodynamics theory.
\newblock {AIAA} Journal \textbf{56}(6), 2478--2483 (2018).
\newblock \doi{10.2514/1.j056400}

\bibitem{Rostami2017}
Rostami, A.B., Armandei, M.: Renewable energy harvesting by vortex-induced
  motions: Review and benchmarking of technologies.
\newblock Renewable and Sustainable Energy Reviews \textbf{70}, 193--214
  (2017).
\newblock \doi{10.1016/j.rser.2016.11.202}

\bibitem{Sclavounos1987}
Sclavounos, P.D.: An unsteady lifting-line theory.
\newblock Journal of Engineering Mathematics \textbf{21}(3), 201--226 (1987).
\newblock \doi{10.1007/bf00127464}

\bibitem{Smyth2019}
Smyth, A.S., Young, A.M., Mare, L.D.: The effect of 3{D} geometry on unsteady
  gust response, using a vortex lattice model.
\newblock In: {AIAA} Scitech 2019 Forum. American Institute of Aeronautics and
  Astronautics (2019).
\newblock \doi{10.2514/6.2019-0899}

\bibitem{Sugar-Gabor2018}
Sugar-Gabor, O.: A general numerical unsteady non-linear lifting line model for
  engineering aerodynamics studies.
\newblock The Aeronautical Journal \textbf{122}(1254), 1199--1228 (2018).
\newblock \doi{10.1017/aer.2018.57}

\bibitem{theodorsen1935}
Theodorsen, T.: General theory of aerodynamic instability and the mechanism of
  flutter.
\newblock Tech. Rep. 496, NACA (1935)

\bibitem{Dyke1964}
{Van Dyke}, M.: Lifting-line theory as a singular-perturbation problem.
\newblock Journal of Applied Mathematics and Mechanics \textbf{28}(1), 90--102
  (1964).
\newblock \doi{10.1016/0021-8928(64)90134-0}

\bibitem{venkata2013leading}
Venkata, S.K., Jones, A.R.: Leading-edge vortex structure over multiple
  revolutions of a rotating wing.
\newblock Journal of aircraft \textbf{50}(4), 1312--1316 (2013)

\bibitem{visbal2017unsteady}
Visbal, M.R.: Unsteady flow structure and loading of a pitching
  low-aspect-ratio wing.
\newblock Physical Review Fluids \textbf{2}(2), 024,703 (2017)

\bibitem{visbal2019dynamic}
Visbal, M.R., Garmann, D.J.: Dynamic stall of a finite-aspect-ratio wing.
\newblock AIAA Journal \textbf{57}(3), 962--977 (2019)

\bibitem{visbal2019effect}
Visbal, M.R., Garmann, D.J.: Effect of sweep on dynamic stall of a pitching
  finite-aspect-ratio wing.
\newblock AIAA Journal \textbf{57}(8), 3274--3289 (2019)

\bibitem{visbal2013three}
Visbal, M.R., Yilmaz, T.O., Rockwell, D.: Three-dimensional vortex formation on
  a heaving low-aspect-ratio wing: computations and experiments.
\newblock Journal of Fluids and Structures \textbf{38}, 58--76 (2013)

\bibitem{Wagner1925}
Wagner, H.: Über die entstehung des dynamischen auftriebes von tragflügeln.
\newblock {ZAMM} - Zeitschrift für Angewandte Mathematik und Mechanik
  \textbf{5}(1), 17--35 (1925).
\newblock \doi{10.1002/zamm.19250050103}

\bibitem{Wang2010}
Wang, Z., Chen, P.C., Liu, D.D., Mook, D.T.:
  Nonlinear-aerodynamics/nonlinear-structure interaction methodology for a
  high-altitude long-endurance wing.
\newblock Journal of Aircraft \textbf{47}(2), 556--566 (2010).
\newblock \doi{10.2514/1.45694}

\bibitem{willis2007combined}
Willis, D.J., Peraire, J., White, J.K.: A combined pfft-multipole tree code,
  unsteady panel method with vortex particle wakes.
\newblock International Journal for Numerical Methods in Fluids \textbf{53}(8),
  1399--1422 (2007)

\bibitem{yilmaz2010scaling}
Yilmaz, T., Ol, M., Rockwell, D.: Scaling of flow separation on a pitching low
  aspect ratio plate.
\newblock Journal of Fluids and Structures \textbf{26}(6), 1034--1041 (2010)

\bibitem{yilmaz2010three}
Yilmaz, T.O., Rockwell, D.: Three-dimensional flow structure on a maneuvering
  wing.
\newblock Experiments in Fluids \textbf{48}(3), 539--544 (2010)

\bibitem{yilmaz2012flow}
Yilmaz, T.O., Rockwell, D.: Flow structure on finite-span wings due to pitch-up
  motion.
\newblock Journal of Fluid Mechanics \textbf{691}, 518--545 (2012)

\end{thebibliography}
